\documentclass[journal]{IEEEtran}
\usepackage[noend]{algpseudocode}
\usepackage{cite}
\usepackage{amsmath,amssymb,amsfonts}
\usepackage{graphicx}
\usepackage{textcomp}
\usepackage{subfigure}
\usepackage{multirow}
\usepackage{color}
\usepackage{diagbox}
\usepackage{algorithm}
\newtheorem{theorem}{Theorem}{}
\newtheorem{lemma}{Lemma}{}
\newtheorem{assumption}{Assumption}
{}
\newtheorem{remark}{Remark}{}
\newtheorem{definition}{Definition}{}
{}
{}
{}
{}

\begin{document}
\title{Data-driven Dynamic Event-triggered Control}
\author{Tao Xu, Zhiyong Sun, Guanghui Wen, and~Zhisheng Duan
\thanks{This work is supported in part by
the National Natural Science Foundation of China under Grants T2121002 and 62173006, and in part by
the China Postdoctoral Science Foundation under Grants 2022TQ0029 and 2022M720435. (Corresponding author: Zhiyong Sun.)}
\thanks{T. Xu is with the School of Mechatronical Engineering,
Beijing Institute of Technology, Beijing, China, and also with the Yangtze Delta Region Academy
of Beijing Institute of Technology, Jiaxing, Zhejiang, China (e-mail: xutao0429@pku.edu.cn).}
\thanks{Z. Sun is with the Department of Electrical Engineering, Eindhoven University of Technology, Eindhoven, the Netherlands (e-mail: sun.zhiyong.cn@gmail.com).}
\thanks{G. Wen is with the Department of Systems Science, Southeast University, Nanjing, China (e-mail: wenguanghui@gmail.com).}
\thanks{Z. Duan is with the Department of Mechanics and Engineering Science, Peking University, Beijing, China (e-mail: duanzs@pku.edu.cn).}
}
\maketitle
\begin{abstract}
  This paper revisits the event-triggered control problem from a data-driven perspective, where unknown continuous-time linear systems subject to disturbances are taken into account.
  Using data information collected off-line instead of accurate system model information, a data-driven dynamic event-triggered control scheme is developed in this paper. The dynamic property is reflected by that the designed event-triggering function embedded in the event-triggering mechanism (ETM) is dynamically updated as a whole.
  Thanks to this dynamic design, a strictly positive minimum inter-event time (MIET) is guaranteed without sacrificing control performance.
  Specifically, exponential input-to-state stability (ISS) of the closed-loop system with respect to disturbances is achieved in this paper, which is superior to some existing results that only guarantee a practical exponential ISS property.
  The dynamic ETM is easy-to-implement in practical operation since all designed parameters
  are determined only by a simple data-driven linear matrix inequality (LMI), without additional complicated conditions as required in relevant literature.
  As quantization is the most common signal constraint in practice, the developed control scheme is further extended to the case where state transmission is affected by a uniform or logarithmic quantization effect.
  Finally, adequate simulations are performed to show the validity and superiority of the proposed control schemes.
\end{abstract}
\section{Introduction}\label{sec1}
\IEEEPARstart{I}{n} ADVANCED control theory and applications, the practical implementation of a controlled system is often based on a digital platform, which mainly consists of physical plants, sensors, samplers, controllers, and actuators. Interestingly,
the sampled information is transmitted to controller over a shared communication network instead of dedicated point-to-point links \cite{Hespanha2007IEEEA, Zhang2016TIISUR}.

The samplers in traditional control setups are time-triggered, in the sense that the sampling instants are determined by continuous time or fixed periods, which will inevitably result in redundant transmissions.
In other words, the communication is artificially forced to be performed either continuously or periodically,
even when it is not actually needed.
As emphasized in the surveys \cite{Hespanha2007IEEEA, Zhang2016TIISUR},
communication networks can only carry a finite amount of information, which restricts the operation of controlled systems and hinders the achievement of control tasks.
Some practical systems are often afflicted by communication resource limitations, such as stealth-required unmanned air vehicles, power-starved planetary rovers, energy-limited underwater vehicles, etc. \cite{Hespanha2007IEEEA}.

To successfully accomplish control tasks without overusing limited communication resources,
event-triggered sampling approach was developed in \cite{Karl1999IFACA} and immediately gained widespread popularity; see some surveys \cite{Nowzari2019AUTOEVE, Ge2020SMCDYN} and references therein.
The idea of event-triggered sampling approach is to transmit information to the network only when it is really needed, which leads to intermittent communication.
Thanks to this resource-aware sampling approach, the sampling instants are no longer passively or artificially generated, but actively and naturally determined by an ETM.
An ETM mainly consists of an event-triggering rule and an event-triggering function, which is the key to balancing resource utilization and control performance.
To date, much attention has been paid on how to design appropriate ETMs, resulting in various static event-triggering functions with state/output-based terms \cite{Abdelrahim2016tacstabilization},
fixed constants \cite{Lunze2010AUTOA}, exponential decay terms \cite{Zhang2014AUTOEVE}, and prescribed timers \cite{Tallapragada2014tacdecentralized}.
To further lower down the triggering frequency, a dynamic event-triggering function composed of a dynamically updated internal variable was built in \cite{Girard2015TACDYN}.
It should be pointed out that Zeno-freeness is an essential property of ETMs, which guarantees the feasibility of event-triggered control schemes.
Since physical devices can only operate at some maximum frequency, ensuring the existence of a strictly positive MIET is more suitable for practical implementation than simply ruling out Zeno phenomenon \cite{Nowzari2019AUTOEVE}.

It is noted that the vast majority of the existing event-triggered control schemes
are built upon the model-driven control, in the sense that explicit knowledge of
system dynamics is indispensable for event-based controller and/or ETM design and implementation.
Obviously, model-driven control is
not applicable when an accurate enough system identification is not available.
In addition, carrying out system identification is expensive and time-consuming, especially for large-scale systems with complicated dynamics.
Recently, data-driven control has attracted considerable attention; see some representative results \cite{Persis2020TACFOR, Berberich2021TACDATA, Bisoffi2022AUTODAT, Joscha2023TACROB, Liu2023TACDATA, Mohammad2023TACDAT} and references therein. This control approach allows control schemes to be developed by state/output and input data without identifying system models.
Recent results on data-driven control subject to event-triggered sampling include \cite{Cordovil2022IJRNCLEARNING, Digge2022ECCDATA, Wang2022arXivDATA} dedicated to discrete-time systems.
For continuous-time systems with linear dynamics in the presence of disturbances, data-driven event-triggered control schemes were developed in
\cite{Wei2023IJRNCDATA, Qi2023TIEDATA, Wang2021arXivDATA, Persis2023TACEVENT}.
Specifically, a static ETM was proposed in \cite{Wei2023IJRNCDATA}, where only a state-based term is designed in the event-triggering function.
Dynamic ETMs with a prescribed timer or a dynamic internal variable were proposed in \cite{Qi2023TIEDATA} and \cite{Wang2021arXivDATA}, respectively.
Note that the setting of the timer in \cite{Qi2023TIEDATA} depends not only on an LMI, but also on some other inequalities, and the gain matrix designed in the ETM in \cite{Wang2021arXivDATA} is determined by a couple of intricate LMIs.
A dynamic ETM consisting of a prescribed timer and a dynamic internal variable was developed in \cite{Persis2023TACEVENT}, which guarantees the existence of a strictly positive MIET at the price of a practical stability property.

In this paper, a data-driven dynamic event-triggered
control scheme is proposed for unknown continuous-time linear systems subject to disturbances.
The major contribution of this paper is to propose a novel ETM with the following distinguished features:
\begin{description}
  \item[(i)] It is easy-to-implement, since the requirement of a prescribed timer as designed in \cite{Qi2023TIEDATA, Persis2023TACEVENT} is removed, and
  the solution of the designed parameters relies only on a simple data-driven LMI, not on an LMI as well as some
       inequalities as in \cite{Qi2023TIEDATA}, nor on a couple of complex LMIs as in \cite{Wang2021arXivDATA}.
  \item[(ii)] It is performance-guaranteed, in the sense that exponential ISS with respect to disturbances is ensured, in contrast to the practical exponential ISS as achieved in \cite{Persis2023TACEVENT}.
  \item[(iii)] It is Zeno-free, as it guarantees the existence of a strictly positive MIET for all time, even in the presence of disturbances.
\end{description}
\noindent As emphasized in \cite{Hespanha2007IEEEA, Zhang2016TIISUR},
the sampled information from controlled systems is typically subject to quantization effects before being transmitted to communication networks.
However, this practical issue is often ignored in the existing literature on data-driven event-triggered control.
This paper further considers that the sampled state information is quantized by uniform quantizers or logarithmic quantizers.
With respect to various quantization scenarios, a quantized version of the developed data-driven dynamic event-triggered control scheme is presented, which is another contribution of this paper.

The remainder of this paper is organized as follows. Preliminaries are introduced in Section~\ref{sec2}. Main results are developed in Section~\ref{sec3}. Some comparisons are provided in Section~\ref{sec4}.
The extensions to quantized scenarios are presented in Section~\ref{sec5}.
Numerical simulations are performed in Section~\ref{sec6}.
 Finally, Section~\ref{sec7} concludes this paper, and appendices contain the proofs.
\section{Preliminaries}\label{sec2}
\subsection{Notation}\label{sec2.1}
We denote the sets of nature number, real number, and non-negative real number by $\mathbb{N}$, $\mathbb{R}$, and $\mathbb{R}_{\geq0}$, respectively.
The symbols $\mathbb{R}^{p}$ and $\mathbb{R}^{p\times q}$ stand for the
$p\times 1$ real vectors and $p\times q$ real matrices, respectively.
The notations $\mathbf{0}$ and $\mathbf{I}$ represent the zero
matrix and the identity matrix, respectively, whose dimensions depend on the context.
Given a real symmetric matrix $A$, the notations
$A\succ0$ ($A\succeq0$) and $A\prec0$ ($A\preceq0$) mean that $A$ is positive definite (semi-definite) and negative definite (semi-definite), respectively.
Denote $\lambda_{m}(A)$ ($\lambda_{M}(A)$) as the minimum (maximum) eigenvalue of a real symmetric matrix $A$.
The notation $\star$ stands for the element of a symmetric block matrix that can be
inferred from symmetry. The symbol $\|\cdot\|$ represents the Euclidean vector norm for a vector and
the induced matrix 2-norm for a matrix.
For $f: \mathbb{R}_{\geq0}\rightarrow\mathbb{R}$, $\|f\|_{[0, t]}$ represents the supremum of $f$ on $[0, t]$, and $f(t^{+})$ stands for the right-hand limit of $f$ at $t$, where $t\geq0$.
For $a\in\mathbb{R}$, $\lfloor a\rceil$ denotes the nearest integer to $a$, where $\lfloor \frac{1}{2}\rceil=1$. The symbol $\mathrm{sign}(\cdot)$ represents the signum function.
\subsection{Problem formulation}\label{sec2.2}
Consider the continuous-time linear system subject to disturbance
\begin{align}\label{eq1}
\dot{x}(t)=Ax(t)+Bu(t)+d(t),
\end{align}
where $A\in\mathbb{R}^{n\times n}$ and $B\in\mathbb{R}^{n\times m}$ are system matrices, and $x(t)\in\mathbb{R}^{n}$, $u(t)\in\mathbb{R}^{m}$ and $d(t)\in\mathbb{R}^{n}$ represent the system state, control input, and disturbance, respectively.

Two common assumptions about system matrices and disturbances are made as follows.
\begin{assumption}\label{as1}
The system matrices $A$ and $B$ are real, constant and unknown, and the pair $(A, B)$ is stabilizable.
\end{assumption}
\begin{assumption}\label{as2}
The disturbance $d(t)$ is unknown, piecewise continuous, and bounded in the sense that $\|d(t)\|\leq\bar{d}<+\infty$.
\end{assumption}

Motivated by \cite{Persis2020TACFOR}, an off-line collected data set is introduced in the following:
\begin{align}\label{eq2}
\mathcal{S}=\{x(t), \dot{x}(t), u(t): t\in\{0, \varsigma, \ldots, (\tau-1)\varsigma\}\},
\end{align}
where $\varsigma$ represents the sampling period and $\tau$ is the number of samplings.
It should be pointed out that the periodical sampling performed in (\ref{eq2}) is to ease the exposition, while aperiodical sampling is also allowed to collect data.
Using the collected data from $\mathcal{S}$, some matrices with data-driven components are defined as follows:
\begin{subequations}
\begin{align}
\mathcal{X}_{0}=&\left[x(0)\ \ x(\varsigma)\ \ \cdots\ \ x\left((\tau-1)\varsigma\right)\right]\in\mathbb{R}^{n\times \tau},\label{eq3a}\\
\mathcal{X}_{1}=&\left[\dot{x}(0)\ \ \dot{x}(\varsigma)\ \ \cdots\ \  \dot{x}\left((\tau-1)\varsigma\right)\right]\in\mathbb{R}^{n\times\tau},\label{eq3b}\\
\mathcal{U}_{0}=&\left[u(0)\ \ u(\varsigma)\ \ \cdots\ \ u\left((\tau-1)\varsigma\right)\right]\in\mathbb{R}^{m\times\tau},\label{eq3c}\\
\mathcal{D}_{0}=&\left[d(0)\ \ d(\varsigma)\ \ \cdots\ \ d\left((\tau-1)\varsigma\right)\right]\in\mathbb{R}^{n\times\tau},\label{eq3d}
\end{align}
\end{subequations}
where $\mathcal{X}_{0}$, $\mathcal{X}_{1}$ and $\mathcal{U}_{0}$ are available, and $\mathcal{D}_{0}$ is unknown.

A couple of assumptions imposed on $\mathcal{X}_{0}$, $\mathcal{U}_{0}$ and $\mathcal{D}_{0}$ are introduced as follows.
\begin{assumption}\label{as3}
The matrix
$[\mathcal{U}_{0}^{T}\ \ \mathcal{X}_{0}^{T}]^{T}$ has full row rank.
\end{assumption}
\begin{assumption}\label{as4}
$\mathcal{D}_{0}\in\mathrm{D}$, where
$\mathrm{D}=\{\mathcal{D}\in\mathbb{R}^{n\times\tau}: \mathcal{D}\mathcal{D}^{T}\preceq\Delta\Delta^{T}\}$,
with
$\Delta\in\mathbb{R}^{n\times n}$ being a known and bounded matrix.
\end{assumption}

The control objective of this paper is to develop data-driven control schemes with dynamic ETMs for continuous-time disturbed linear systems (\ref{eq1}) without degrading system performance, in the absence and presence of quantization.
\section{Main results}\label{sec3}
In this section, a data-driven dynamic event-triggered control scheme is developed, and some necessary discussions are provided.

The event sequence is denoted as $\{t_{k}\}_{k\in\mathbb{N}}$, with $t_{0}=0$, which is determined by
the following event-triggering rule:
\begin{align}\label{eq4}
t_{k+1}=\inf\left\{t>t_{k}|\ f(t)\leq0\right\},
\end{align}
where $f(t)$ represents the dynamic event-triggering function.
For $t\in[t_{k}, t_{k+1})$, the update algorithm of $f(t)$ is proposed as
\begin{align}\label{eq5}
\dot{f}(t)=\min\left\{-z(t)^{T}\Phi z(t), 0\right\}-f(t), f(t_{k}^{+})=\bar{f},
\end{align}
where the constant $\bar{f}>0$, and $z(t)=[x(t)^{T}\ \ e(t)^{T}]^{T}$ with
\begin{align}\label{eq6}
e(t)=x(t_{k})-x(t),
\end{align}
and
\begin{align}\label{eq7}
\Phi=\left[\begin{array}{cc}
-\alpha\mathbf{I}&\mathbf{0}\\
\mathbf{0}&\beta\mathbf{I}\\
\end{array}\right],
\end{align}
where the constants $\alpha, \beta>0$ are determined by a data-driven LMI to be proposed later.
Once an event is triggered by (\ref{eq4}), $e(t)$ will be $0$ and $f(t)$ will be instantly reset to $\bar{f}$.
It is easy to verify that $\dot{f}(t)\leq0$ and $0\leq f(t)\leq\bar{f}$.

Using the state information sampled at event instants, the event-based controller is proposed as
\begin{align}\label{eq8}
u(t)=Kx(t_{k}), t\in[t_{k}, t_{k+1}),
\end{align}
where $K\in\mathbb{R}^{m\times n}$ represents the data-driven feedback gain matrix designed off-line, of the form
\begin{align}\label{eq9}
K=\mathcal{U}_{0}\mathcal{Y}(\mathcal{X}_{0}\mathcal{Y})^{-1},
\end{align}
where the matrix $\mathcal{Y}\in\mathbb{R}^{\tau\times n}$ is a solution of
\begin{align}
\left[\begin{array}{cc}
\mathcal{X}_{1}\mathcal{Y}+(\mathcal{X}_{1}\mathcal{Y})^{T}+\Omega+\gamma\Delta\Delta^{T}&\mathcal{Y}^{T}\\
\mathcal{Y}&-\gamma\mathbf{I}\\
\end{array}\right]\prec&0,\label{eq10}\\
\mathcal{X}_{0}\mathcal{Y}\succ&0,\label{eq11}
\end{align}
where $\gamma>0$ is a proper constant, and $\Omega\succ0$ is a given matrix.

Let $\mathcal{P}=(\mathcal{X}_{0}\mathcal{Y})^{-1}$, which is positive definite under (\ref{eq11}), and let $\mathcal{Q}\in\mathbb{R}^{\tau\times n}$ be any solution of
\begin{align}\label{eq12}
\left[\begin{array}{c}
K\\
\mathbf{0}\\
\end{array}\right]=
\left[\begin{array}{c}
\mathcal{U}_{0}\\
\mathcal{X}_{0}\\
\end{array}\right]\mathcal{Q},
\end{align}
which exists under Assumption~\ref{as3}.

Before presenting the main theorem, a definition relevant to system stability is introduced in the following:
\begin{definition}\label{de1}
Suppose that there exist constants $c_{1}\geq1$, $c_{2}, c_{3}>0$, and $c_{4}\geq0$, such that
any solution $x(t)$ to (\ref{eq1}) with input disturbance $d(t)$ in closed-loop with (\ref{eq8})
satisfies
\begin{align}\label{eq13}
\|x(t)\|\leq c_{1}e^{-c_{2}t}\|x(0)\|+c_{3}\|d(t)\|_{[0, t]}+c_{4}, \forall t\geq0.
\end{align}
Then, the system (\ref{eq1}) is with exponential ISS property with respect to disturbance if $c_{4}=0$, and the system (\ref{eq1}) is with practical exponential ISS property if $c_{4}>0$.
\end{definition}

Now, we are ready to present the main results of this section.
\begin{theorem}\label{th1}
Suppose that Assumptions~\ref{as1}--~\ref{as4} hold, and there exist proper positive constants $\alpha, \beta, \delta$, such that
the following data-driven LMI holds
\begin{align}\label{eq14}
\left[\begin{array}{ccc}
-\frac{\delta}{8}\mathcal{P}\Omega\mathcal{P}+\alpha\mathbf{I}&\delta\mathcal{P}\mathcal{X}_{1}\mathcal{Q}&\delta\mathcal{P}\Delta\\
\star&\gamma\mathcal{Q}^{T}\mathcal{Q}-\beta\mathbf{I}&\mathbf{0}\\
\star&\star&-\gamma\mathbf{I}\\
\end{array}\right]\preceq0.
\end{align}
Using the dynamic ETM (\ref{eq4})--(\ref{eq7}), the event-based controller (\ref{eq8}), and the data-driven feedback gain matrix (\ref{eq9}), then,
\begin{description}
  \item[(i)] the system (\ref{eq1}) features the exponential ISS property with respect to disturbance;
  \item[(ii)] the inter-event intervals are lower bounded by a strictly positive constant.
\end{description}
\end{theorem}

The proof of Theorem~\ref{th1} is presented in Appendix~\ref{app1}.

\begin{remark}\label{re2}
(How to obtain data-driven matrix $\mathcal{X}_{1}$ in practice?)
It is difficult to obtain $\mathcal{X}_{1}$ in practice since the computation of $\dot{x}(t)$ is error-prone.
In the absence of disturbances, an Euler discretization-based approach is proposed in \cite{Berberich2021arXivdata}, where
the $i$th component of $\mathcal{X}_{1}$, $\dot{x}(i\varsigma)$, $i=0, 1, \ldots, \tau-1$, is approximated by $\hat{\dot{x}}(i\varsigma)=\frac{x((i+1)\varsigma)-x(i\varsigma)}{\varsigma}$ with the approximation error
satisfying $\|\hat{\dot{x}}(i\varsigma)-\dot{x}(i\varsigma)\|\leq\frac{\bar{a}\varsigma}{2}[\bar{a}\|x(i\varsigma)\|
+(1+\frac{\bar{a}\varsigma}{3})\bar{b}\|u(i\varsigma)\|]$, where $\bar{a}$ and $\bar{b}$ represent the known upper bounds of $\|A\|$ and $\|B\|$, respectively.
When disturbances are considered, the authors in \cite{Persis2023TACEVENT} present an integral-based approach, where a new data set that does not involve the computation of $\dot{x}(t)$ is constructed by integrating both sides of $\dot{x}(t)=Ax(t)+Bu(t)+d(t)$ and performing re-sampling.
\end{remark}
\begin{remark}\label{re3}
(How to implement the developed control scheme?)
To clearly show the implementation process of the developed data-driven dynamic event-triggered control scheme, a block diagram is presented in Fig.~\ref{fig1} and an algorithm is provided in Algorithm~\ref{alg1}.
In can be observed from Fig.~\ref{fig1} that event generator, data processor, and zero-order holder are some essential devices.
More specifically, event generator monitors the dynamic ETM and generates events on-line.
Data processor collects data information and calculates the data-driven feedback gain matrix $K$, which are all performed off-line. Zero-order holder keeps $u(t)=Kx(t_{k})$ unchanged for $t\in[t_{k}, t_{k+1})$ until the next event is triggered.
\begin{figure}[!t]
\centering
{\includegraphics[width=3in]{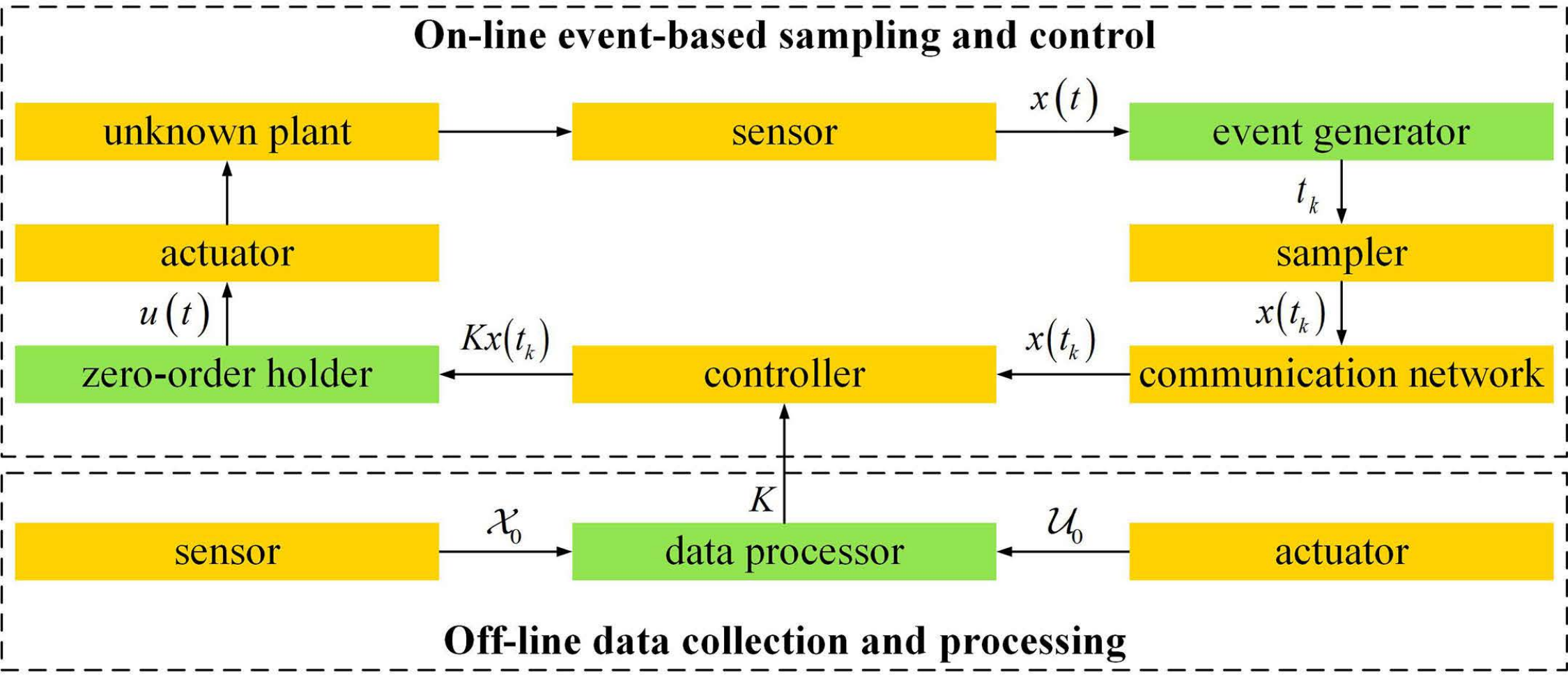}}\hspace{5pt}
\caption{Implementation process of the proposed data-driven dynamic event-triggered control scheme, where the devices in the orange blocks are common in traditional control mode, and those in the green blocks are essential in the data-driven event-triggered control framework.} \label{fig1}
\end{figure}
\begin{algorithm}[!t]
\caption{Data-driven dynamic event-triggered control}\label{alg1}
\textbf{Off-line operation}
\begin{algorithmic}[1]
\State Collect data information $\mathcal{X}_{0}$, $\mathcal{U}_{0}$, and $\mathcal{X}_{1}$.
\State Calculate the feedback gain matrix $K$ proposed in (\ref{eq9}).
\end{algorithmic}
\textbf{On-line operation}
\begin{algorithmic}[1]
\State Construct the measurement error $e(t)$ defined in (\ref{eq6}).
\State Choose $\alpha$ and $\beta$ according to the LMI given in (\ref{eq14}).
\State Update the event-triggering function $f(t)$ based on (\ref{eq5}).
\If {$f(t)\leq0$}
\State Generate $t_{k}$, sample and broadcast $x(t_{k})$, and reset $f(t)=\bar{f}$.
\EndIf
\State Operate the controller $u(t)$ developed in (\ref{eq8}).
\end{algorithmic}
\end{algorithm}
\end{remark}
\begin{remark}\label{re4}
(How to increase the inter-event intervals?)
From a qualitative perspective, the event-triggering function $f(t)$ is a decay function,
and a larger $\bar{f}$ will increase the value at which $f(t)$ begins to decay and a smaller $\beta$ will decrease the decay rate of $f(t)$.
From a quantitative perspective, the explicit expression of the lower bound of inter-event intervals is given in (\ref{eq44}), indicating that a larger $\bar{f}$ or a smaller $\beta$ will result in a larger value.
Therefore, the inter-event intervals can be increased by choosing a larger $\bar{f}$ or a smaller $\beta$.
\end{remark}
\section{Comparisons}\label{sec4}
In this section, some existing dynamic ETMs for data-driven event-triggered control are briefly reviewed for comparisons, where some notations are changed to facilitate representation.
\subsection{Dynamic ETMs with a prescribed timer or a dynamic internal variable}\label{sec4.1}
The dynamic ETM proposed in \cite{Qi2023TIEDATA} is as follows:
\begin{align}\label{eq15}
t_{k+1}=\inf\left\{t>t_{k}+\tilde{\rho}|\ \tilde{\eta}(t)\leq0\right\},
\end{align}
where the update law of $\tilde{\eta}(t)$ can be referred to \cite[(19)]{Qi2023TIEDATA} and the setting of the timer $\tilde{\rho}$ depends not only on an LMI \cite[(21)]{Qi2023TIEDATA}, but also on some inequalities, e.g., \cite[(22) and (23)]{Qi2023TIEDATA}.

The dynamic ETM proposed in \cite{Wang2021arXivDATA} is presented in the following:
\begin{align}\label{eq16}
t_{k+1}=t_{k}+h\min\left\{j>0|\ \tilde{\eta}(\mu_{k}^{j})+\tilde{\vartheta}\tilde{\rho}(\mu_{k}^{j})<0\right\},
\end{align}
where $\mu_{k}^{j}=t_{k}+jh$, $j=0, 1, \ldots$, with $h$ being the prescribed sampling period, the constant $\tilde{\vartheta}$ is specified in \cite[Theorem~1]{Wang2021arXivDATA}, and
the explicit expression of $\tilde{\rho}(\mu_{k}^{j})$ and the update law of the dynamic internal variable $\tilde{\eta}(\mu_{k}^{j})$ can be found in \cite[(12)]{Wang2021arXivDATA} and \cite[(13)]{Wang2021arXivDATA}, respectively.
Moreover,
the choice of the gain matrix designed in \cite[(12)]{Wang2021arXivDATA} is determined by a couple of LMIs \cite[(18) and (19)]{Wang2021arXivDATA} with complex components detailed in \cite[Theorem~1]{Wang2021arXivDATA}.

Compared with the dynamic ETMs proposed in \cite{Qi2023TIEDATA} and \cite{Wang2021arXivDATA}, less conservative conditions are required to implement the dynamic ETM proposed in (\ref{eq4})--(\ref{eq7}). Specifically, the design of a prescribed timer is removed and the designed parameters, e.g., $\alpha, \beta$, are determined only by a data-driven LMI (\ref{eq14}) whose components are much simpler than those in \cite{Wang2021arXivDATA}.
\subsection{A dynamic ETM with a prescribed timer and a dynamic internal variable}\label{sec4.2}
A dynamic ETM with a prescribed timer $\bar{\rho}$ as well as an interval variable $\eta(t)$ is designed in \cite{Persis2023TACEVENT}, as shown below:
\begin{align}\label{eq17}
t_{k+1}=\inf\left\{t\geq t_{k}+\bar{\rho}|\ \vartheta(z(t)^{T}\Psi z(t)-\nu)-\eta(t)\geq0\right\},
\end{align}
where the constants $\nu, \vartheta\geq0$, the matrix $\Psi$ can be set as \cite[(28)]{Persis2023TACEVENT}, the timer $\bar{\rho}\in\{0, \rho\}$, with $\rho>0$ defined in \cite[(41)]{Persis2023TACEVENT}, which is dependent on $\mathcal{X}_{0}$, $\mathcal{X}_{1}$ and $\mathcal{U}_{0}$.
Moreover, the interval variable $\eta(t)$ is dynamically updated by \cite[(54)]{Persis2023TACEVENT}.

Note that a strictly positive MIET can be ensured when applying (\ref{eq17}) with $\nu>0$ or $\bar{\rho}=\rho$.
When $\nu>0$, the system (\ref{eq1}) is only with practical exponential ISS property.
When $\bar{\rho}=\rho$, the event-based controller will not update until at least time $\rho$ has elapsed, which will prolong the time for the closed-loop system to achieve stability.
In this paper, the existence of a strictly positive MIET is guaranteed, without requiring a prescribed timer and degrading system stability.
\section{Extensions}\label{sec5}
In this section, we aim to extend the main results developed in Section~\ref{sec3} to the scenario where state information is quantized by various quantizers $q_{\ast}: \mathbb{R}\rightarrow\mathbb{R}$, where $\ast=u$ or $\ast=l$ represents the uniform quantizer or logarithmic quantizer, respectively.

Consider the same $q_{u}$ and $q_{l}$ as in \cite{Guo2013AUTOCON}, which are respectively detailed as follows:
\begin{align}
q_{u}(x_{i}(t))=&\theta\left\lfloor\frac{x_{i}(t)}{\theta}\right\rceil,\label{eq18}\\
q_{l}(x_{i}(t))=&\mathrm{sign}(x_{i}(t))e^{q_{u}[\ln|x_{i}(t)|]},\label{eq19}
\end{align}
where the constant $\theta>0$ represents the quantizer parameter, and $x_{i}(t)\in\mathbb{R}$ represents the $i$th component of $x(t)$.
Let $q_{\ast}(x(t))$ be the column stack vector of $q_{\ast}(x_{i}(t))$, $i=1, \ldots, n$.
Denoted by $\epsilon_{\ast}(t)=q_{\ast}(x(t))-x(t)$ the quantization error. It holds that $\|\epsilon_{u}(t)\|\leq\frac{\sqrt{n}\theta}{2}$ and
$\|\epsilon_{l}(t)\|\leq(e^{\frac{\theta}{2}}-1)\|x(t)\|$.

Under uniform or logarithmic quantization effect, the dynamic ETM developed in (\ref{eq4})--(\ref{eq7}) is modified as
\begin{align}\label{eq20}
t_{k+1}=\inf\left\{t>t_{k}|\ f_{\ast}(t)\leq0\right\},
\end{align}
where the update algorithm of $f_{\ast}(t)$, $t\in[t_{k}, t_{k+1})$, is proposed as
\begin{align}\label{eq21}
\dot{f}_{\ast}(t)=\min\left\{-z_{\ast}(t)^{T}\Phi_{\ast}z_{\ast}(t), 0\right\}-f_{\ast}(t), f_{\ast}(t_{k}^{+})=\bar{f}_{\ast},
\end{align}
where the constant $\bar{f}_{\ast}>0$, $z_{\ast}(t)=[q_{\ast}(x(t))^{T}\ \ e_{\ast}(t)^{T}]^{T}$, with
\begin{align}\label{eq22}
e_{\ast}(t)=q_{\ast}(x(t_{k}))-q_{\ast}(x(t)),
\end{align}
and
\begin{align}\label{eq23}
\Phi_{u}=\left[\begin{array}{cc}
-\frac{\alpha}{2}\mathbf{I}&\mathbf{0}\\
\mathbf{0}&\beta\mathbf{I}\\
\end{array}\right],
\Phi_{l}=\left[\begin{array}{cc}
-e^{-\theta}\alpha\mathbf{I}&\mathbf{0}\\
\mathbf{0}&\beta\mathbf{I}\\
\end{array}\right],
\end{align}
where the constants $\alpha, \beta>0$.

The event-based quantized controller is designed as
\begin{align}\label{eq24}
u(t)=Kq_{\ast}(x(t_{k})), t\in[t_{k}, t_{k+1}),
\end{align}
with the data-driven feedback gain matrix $K$ proposed in (\ref{eq9}).

Definition~\ref{de1} proposed in Section~\ref{sec3} is applied in this section with (\ref{eq8}) replaced by (\ref{eq24}) in the closed-loop system.
The main theorem of this section is provided in the following.
\begin{theorem}\label{th2}
Suppose that Assumptions~\ref{as1}--~\ref{as4} hold, and there exist proper positive constants $\alpha, \beta, \delta$, such that LMI (\ref{eq14}) holds.
Under the dynamic ETM (\ref{eq20})--(\ref{eq23}), the
event-based quantized controller (\ref{eq24}), and the data-driven feedback gain matrix (\ref{eq9}), the following hold:
\begin{description}
  \item[(i)] the system (\ref{eq1}) subject to uniform quantization is with practical exponential ISS;
  \item[(ii)] the system (\ref{eq1}) subject to logarithmic quantization features the exponential ISS property with respect to disturbance;
  \item[(iii)] there exists a strictly positive lower bound of the inter-event intervals in the presences of uniform or logarithmic quantization.
\end{description}
\end{theorem}

The proof of Theorem~\ref{th2} is presented in Appendix~\ref{app2}.
\section{Simulations}\label{sec6}
Some simulation examples are carried out in this section to illustrate the effectiveness of the proposed control schemes.

Consider system (\ref{eq1}) that models the linearized longitudinal dynamics of an aircraft with the following system matrices
\cite{heck1989jgcdapplication}
\begin{align}
A=\left(\begin{array}{ccc}
-0.277&1&-0.0002 \\
-17.1&-0.178&-12.2 \\
0&0&-6.67\\
\end{array}\right),
B=\left(\begin{array}{c}
0\\
0\\
6.67\\
\end{array}\right).\notag
\end{align}
To collect useful data, we run an experiment with input uniformly distributed in $[-1, 1]$ and
initial state uniformly distributed in $[-10, 10]$, and keep the input constant during the intervals $[0, \varsigma], [\varsigma, 2\varsigma], \ldots, [(\tau-2)\varsigma, (\tau-1)\varsigma]$, where
$\varsigma=0.1$ and $\tau=10$, which satisfies that $\tau\geq n+m=4$.
We set $\Delta=\sqrt{\tau}\bar{d}\mathbf{I}$, choose $\Omega=7\mathbf{I}$, solve (\ref{eq10}) and (\ref{eq11}) to obtain $\gamma$ and $\mathcal{Y}$, compute (\ref{eq9}) to derive $K$, and solve (\ref{eq14}) to obtain $\alpha, \beta, \delta$. The solutions of $K, \alpha, \beta, \delta, \gamma$ under $\bar{d}=0.1, 0.2, 0.3$ are provided in Table~\ref{tab1}.
\begin{table*}[!htb]
  \centering
  \caption{The solutions of $K, \alpha, \beta, \delta, \gamma$ under different $\bar{d}$.}
  \label{tab1}
  \tabcolsep=0.2cm
  \renewcommand\arraystretch{1.3}
  \begin{tabular}{c c c c c c}
  \hline
    $\bar{d}$&$K$&$\alpha$&$\beta$&$\delta$&$\gamma$\\
  \hline
    0.1&$[-0.1178\ \ 0.2356\ \ 	0.0866]$&$1.6998\times10^{-2}$&$5.5117\times10^{3}$&$7.9030\times10$&$2.2520\times10$\\
  \hline
    0.2&$[-0.1456\ \ 0.2145\ \	0.0065]$&$3.9422\times10^{-3}$&$2.9814\times10^{4}$&$2.1549\times10$&$2.1961\times10$\\
    \hline
  0.3&$[-0.1827\ \ 0.1886\ \ -0.0760]$&$1.1984\times10^{-3}$&$3.0452\times10^{4}$&$1.0740\times10$&$2.0470\times10$\\
    \hline
  \end{tabular}
\end{table*}
\subsection{Simulation results in the absence of quantization}\label{sec6.1}
\begin{figure*}[!htb]
\centering
\subfigure[$\bar{d}=0.1$.]{
\includegraphics[width=5.5cm]{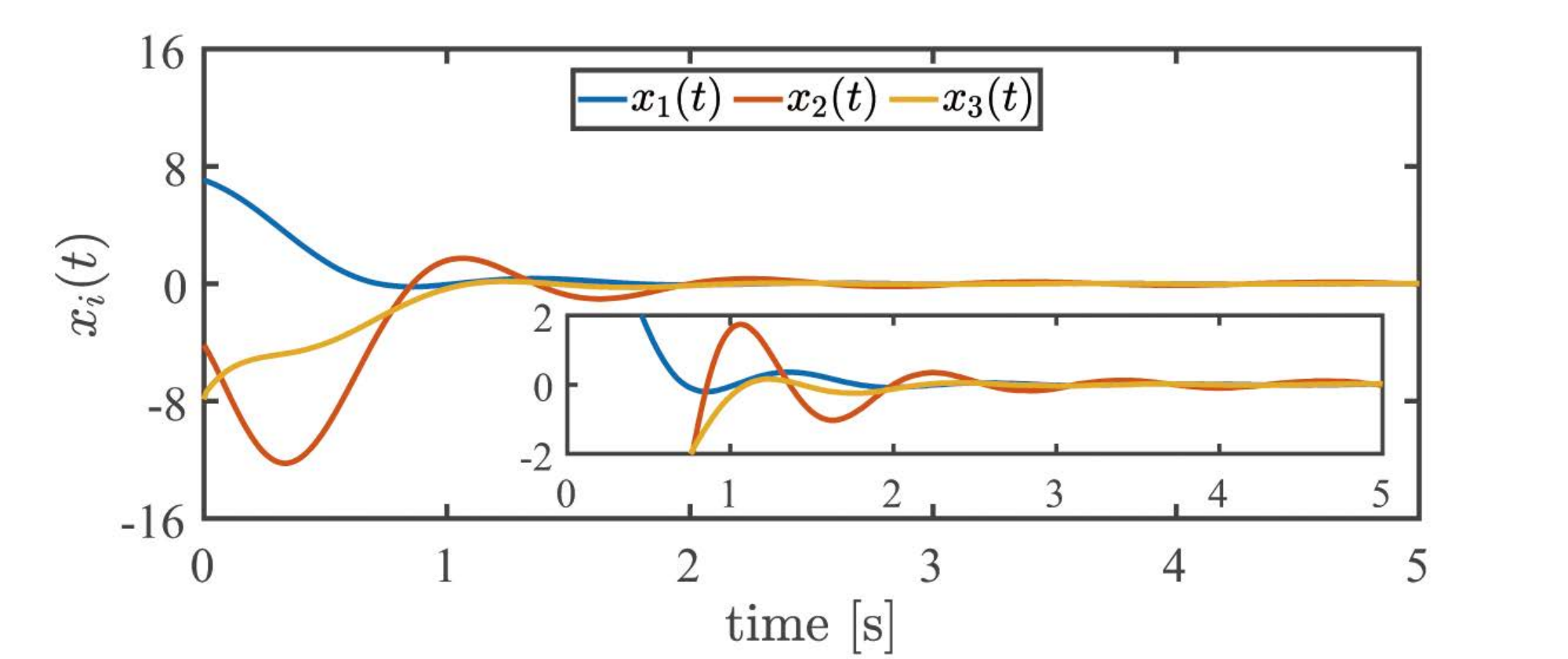}\label{fig2a}
}
\hspace{-9mm}
\subfigure[$\bar{d}=0.2$.]{
\includegraphics[width=5.5cm]{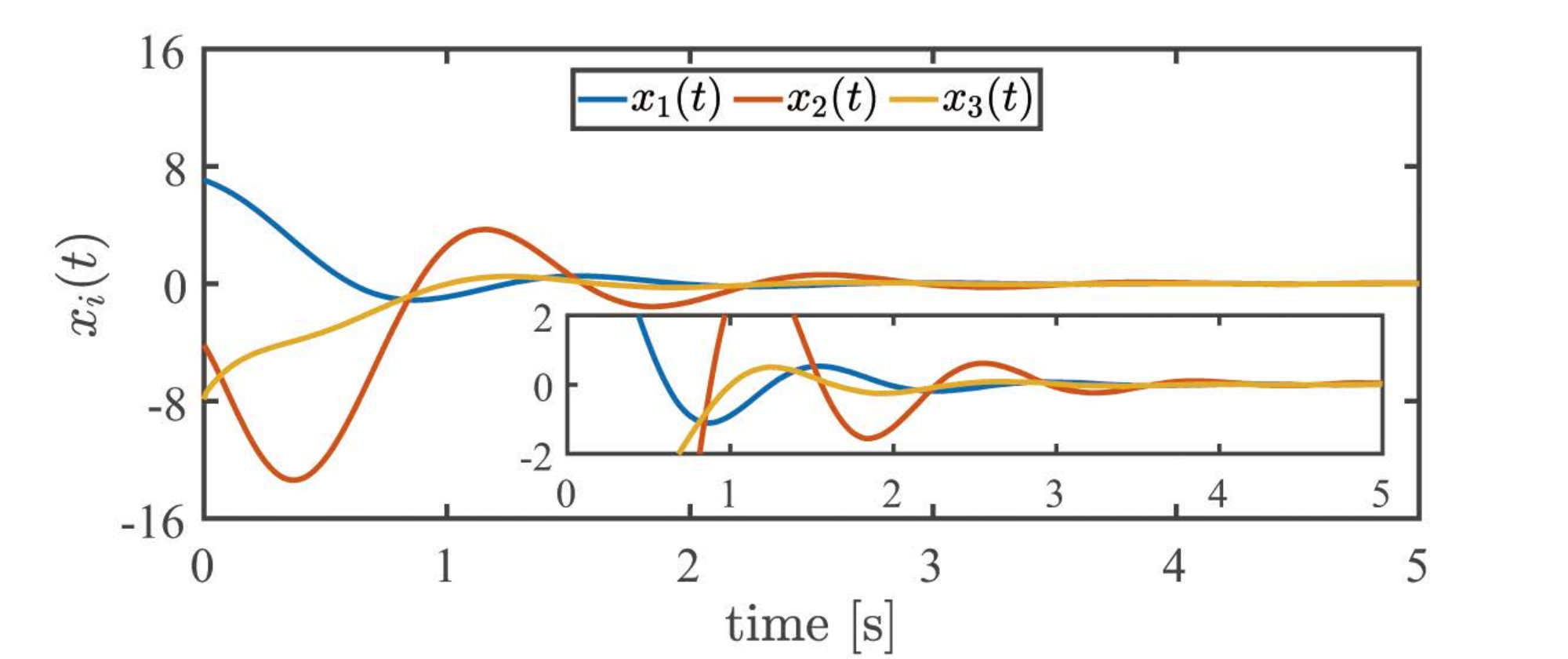}\label{fig2b}
}
\hspace{-9mm}
\subfigure[$\bar{d}=0.3$.]{
\includegraphics[width=5.5cm]{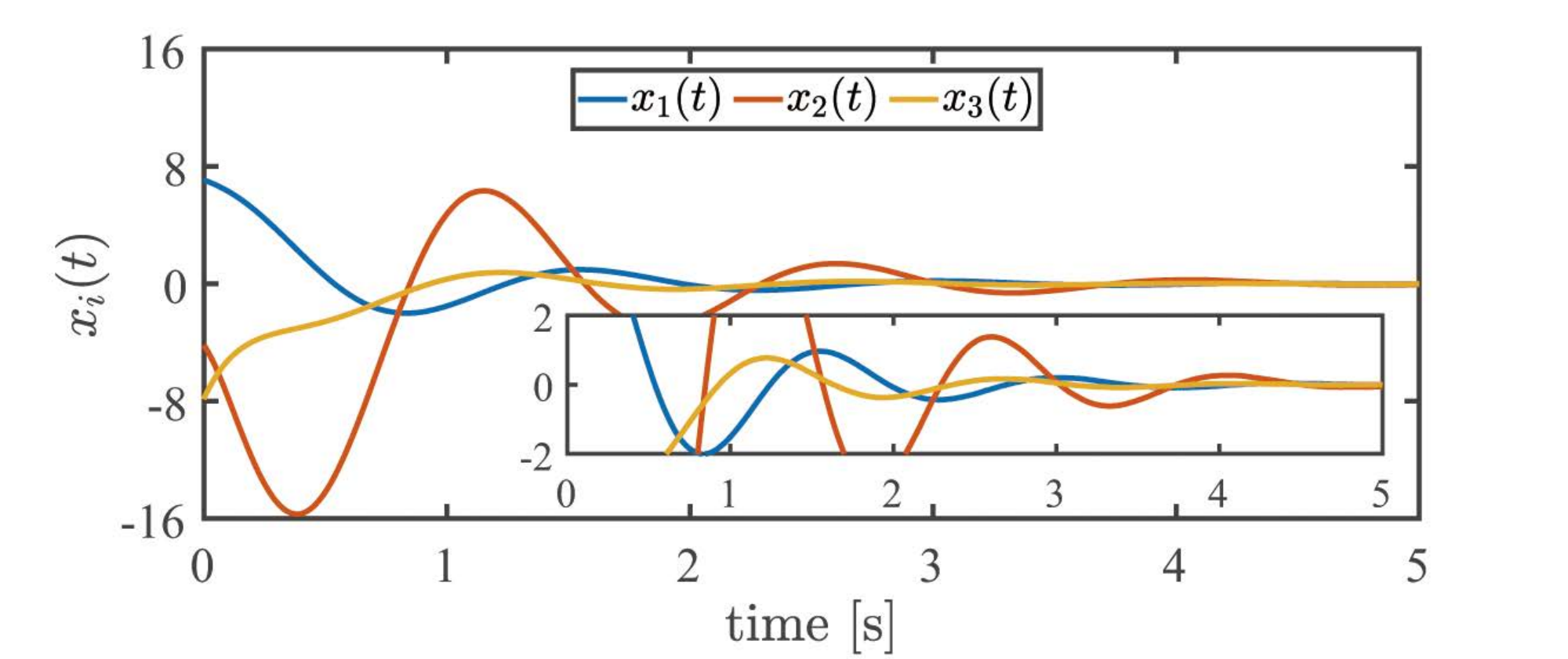}\label{fig2c}
}
\caption{The trajectories of each component of the system state $x(t)$ under fixed $\bar{f}=100$ and different $\bar{d}$.}\label{fig2}
\end{figure*}

\begin{figure*}[!htb]
\centering
\subfigure[$0\mathrm{s}\leq t\leq0.5\mathrm{s}$.]{
\includegraphics[width=5.5cm]{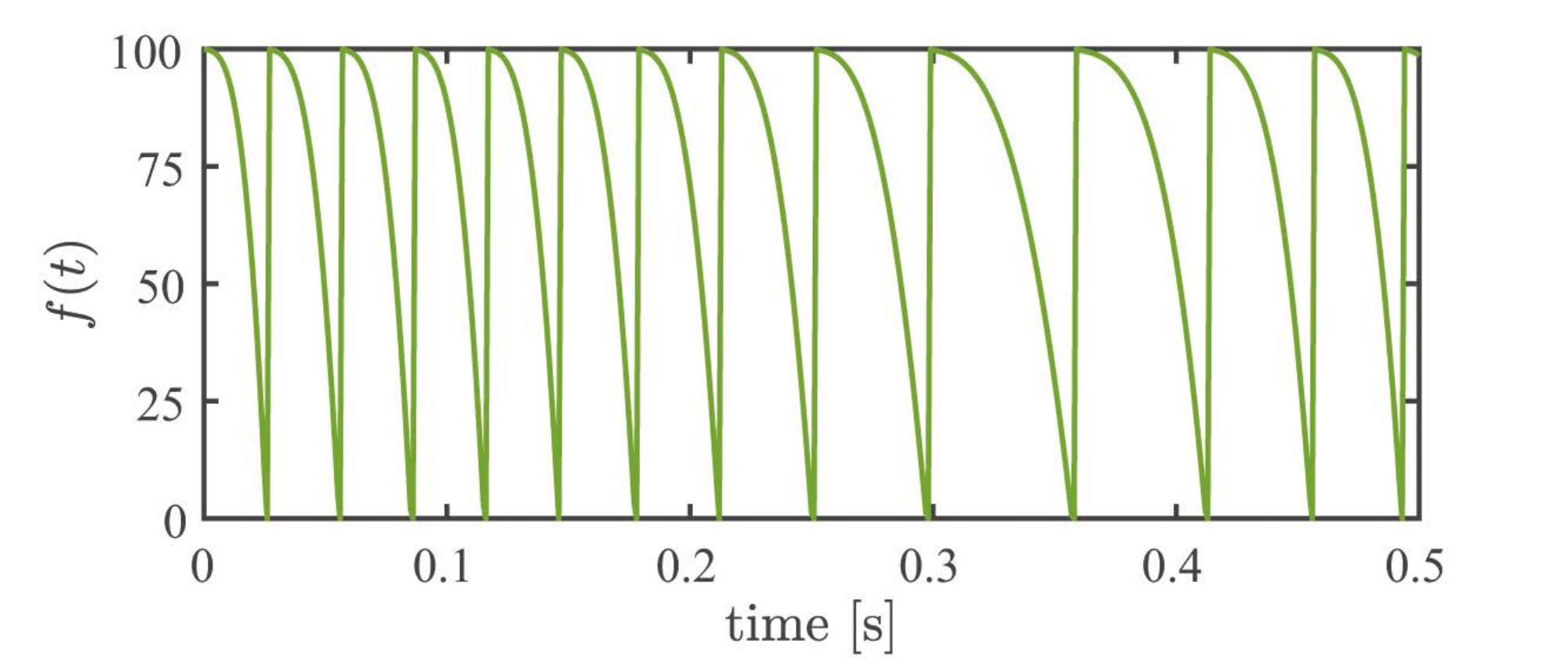}\label{fig3a}
}
\hspace{-9mm}
\subfigure[$0.5\mathrm{s}\leq t\leq1.5\mathrm{s}$.]{
\includegraphics[width=5.5cm]{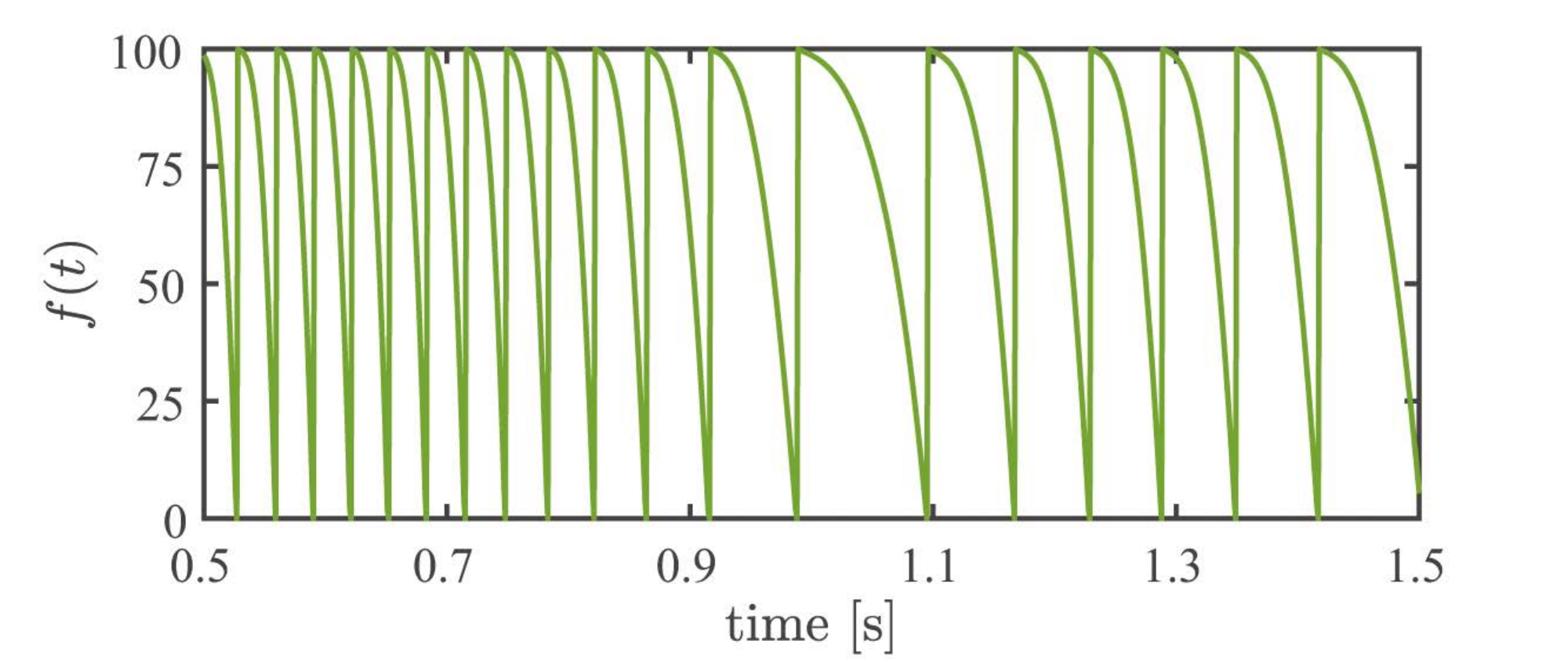}\label{fig3b}
}
\hspace{-9mm}
\subfigure[$1.5\mathrm{s}\leq t\leq5\mathrm{s}$.]{
\includegraphics[width=5.5cm]{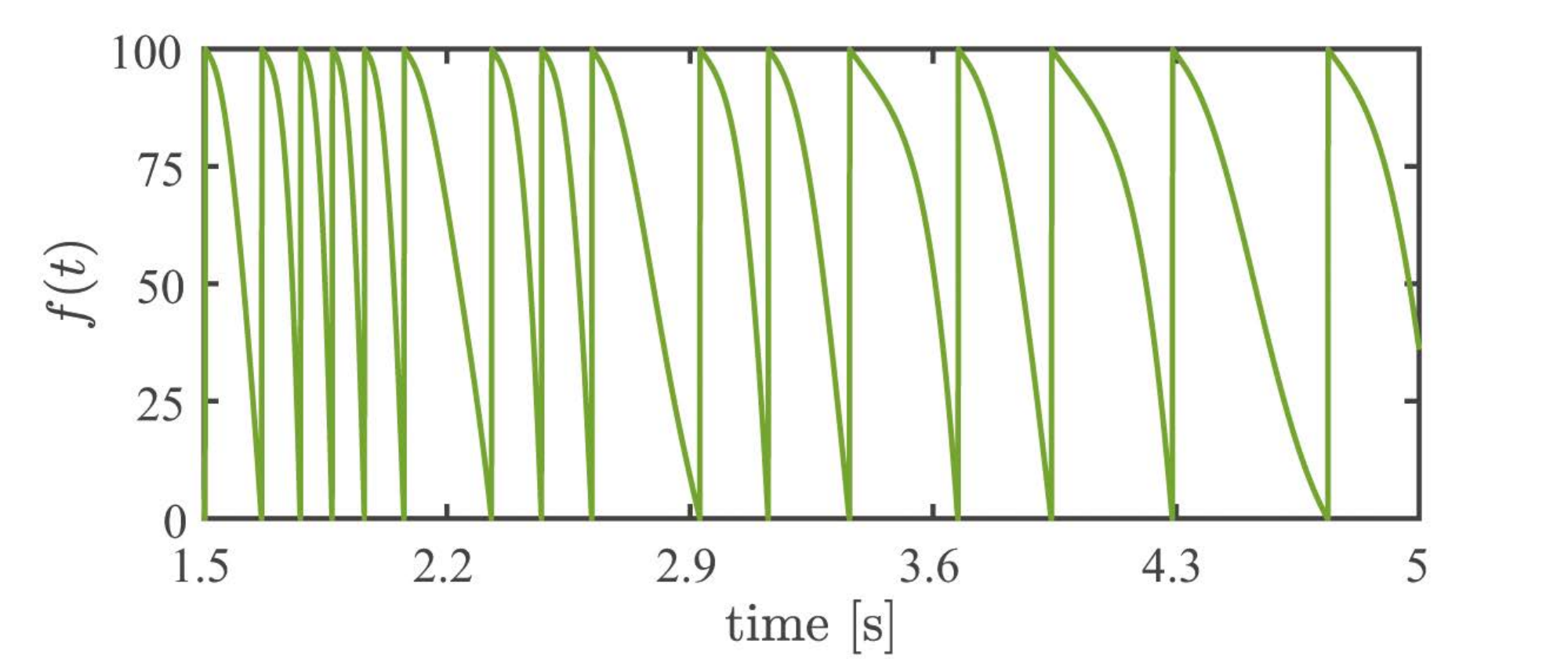}\label{fig3c}
}
\caption{The evolution of the dynamic event-triggering function $f(t)$.}\label{fig3}
\end{figure*}

To show the effectiveness of the data-driven dynamic event-triggered control scheme proposed in Section~\ref{sec3} and the effect of the upper bound of disturbance $\bar{d}$, the evolution of each component of the system state $x(t)$, i.e., $x_{i}(t)$, $i=1, 2, 3$, under fixed $\bar{f}=100$, fixed initial states, and different $\bar{d}$ is exhibited in Fig.~\ref{fig2}. It can be observed from Fig.~\ref{fig2} that each component of $x(t)$ gradually tends to the neighborhood of zero point, and
the deviation between each component of $x(t)$ and zero point increases
as $\bar{d}$ increases.
To illustrate the triggering performance of the dynamic ETM, the trajectory of the dynamic event-triggering function $f(t)$ under $\bar{d}=0.1$ and $\bar{f}=100$ is shown in Fig.~\ref{fig3}. It follows from Fig.~\ref{fig3} that $f(t)$ decays from $\bar{f}$ and instantly resets to $\bar{f}$ when $f(t)$ reaches $0$.
Moreover, it is calculated that
the total number of the event instants during the operation time is $48$, the MIET derived from the simulation is $3.0\times10^{-2}\mathrm{s}$, and the theoretical lower bound derived from (\ref{eq44}) is $8.7\times10^{-3}\mathrm{s}$. It is shown that only a finite number of events are triggered during the operation time, and the MIET is strictly positive and larger than the theoretical lower bound of the inter-event intervals, which is sufficient to confirm the exclusion of Zeno phenomenon.

\subsection{Simulation results in the presence of quantization}\label{sec6.2}
\begin{figure*}[!htb]
\centering
\subfigure[$\theta=0.1$.]{
\includegraphics[width=5.5cm]{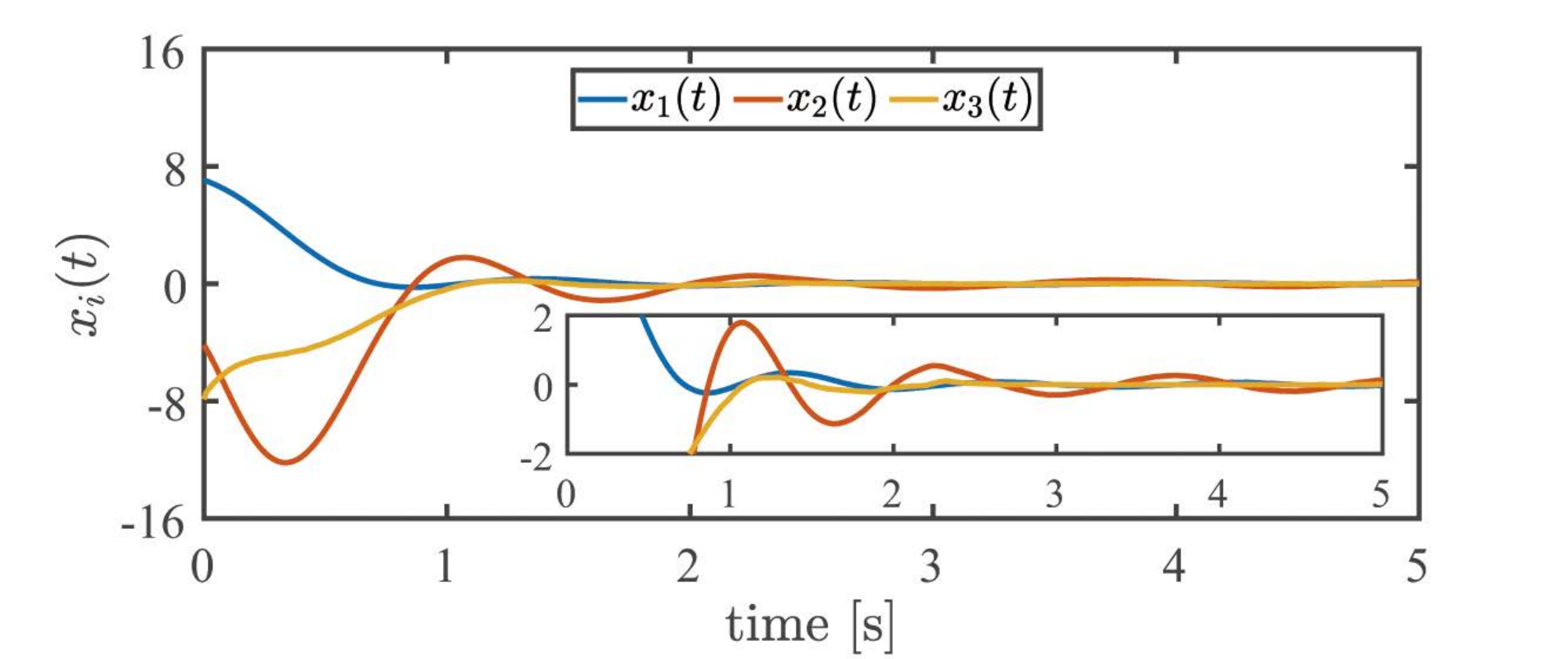}\label{fig4a}
}
\hspace{-9mm}
\subfigure[$\theta=0.2$.]{
\includegraphics[width=5.5cm]{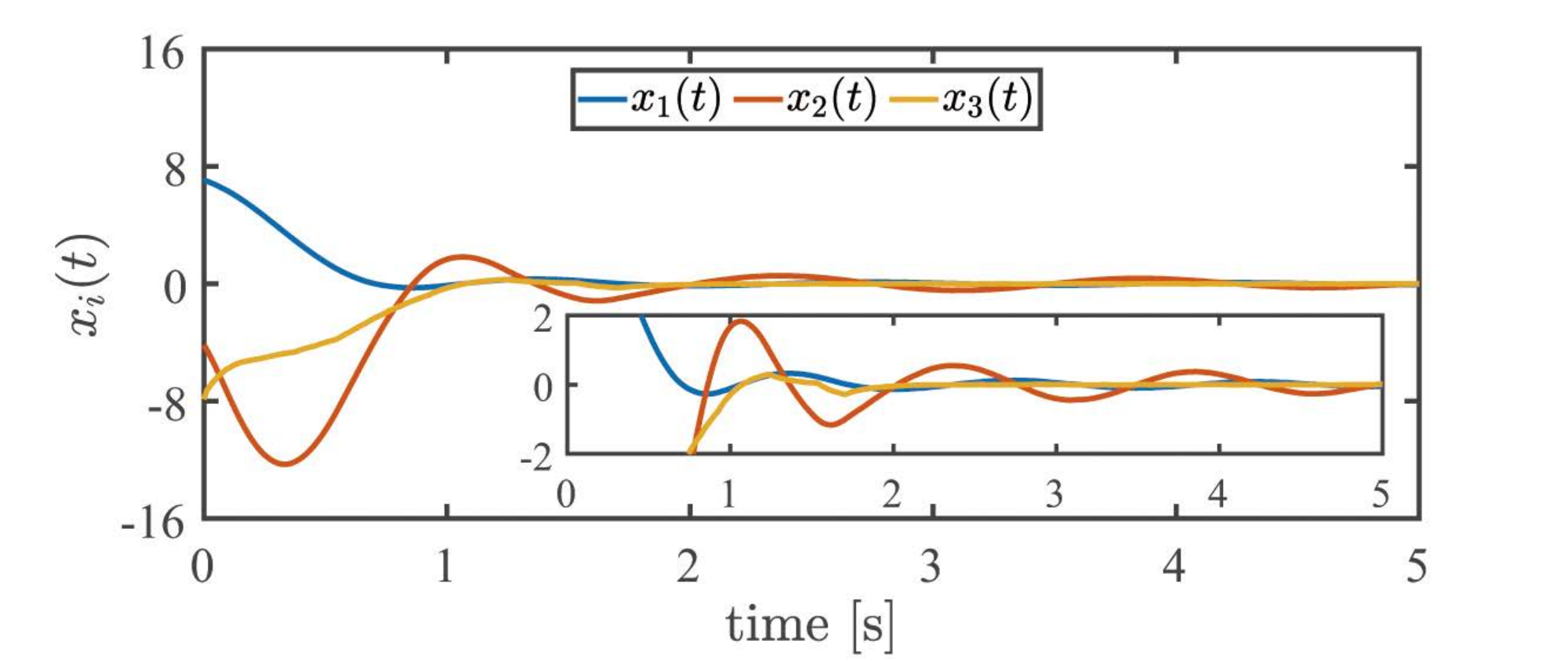}\label{fig4b}
}
\hspace{-9mm}
\subfigure[$\theta=0.3$.]{
\includegraphics[width=5.5cm]{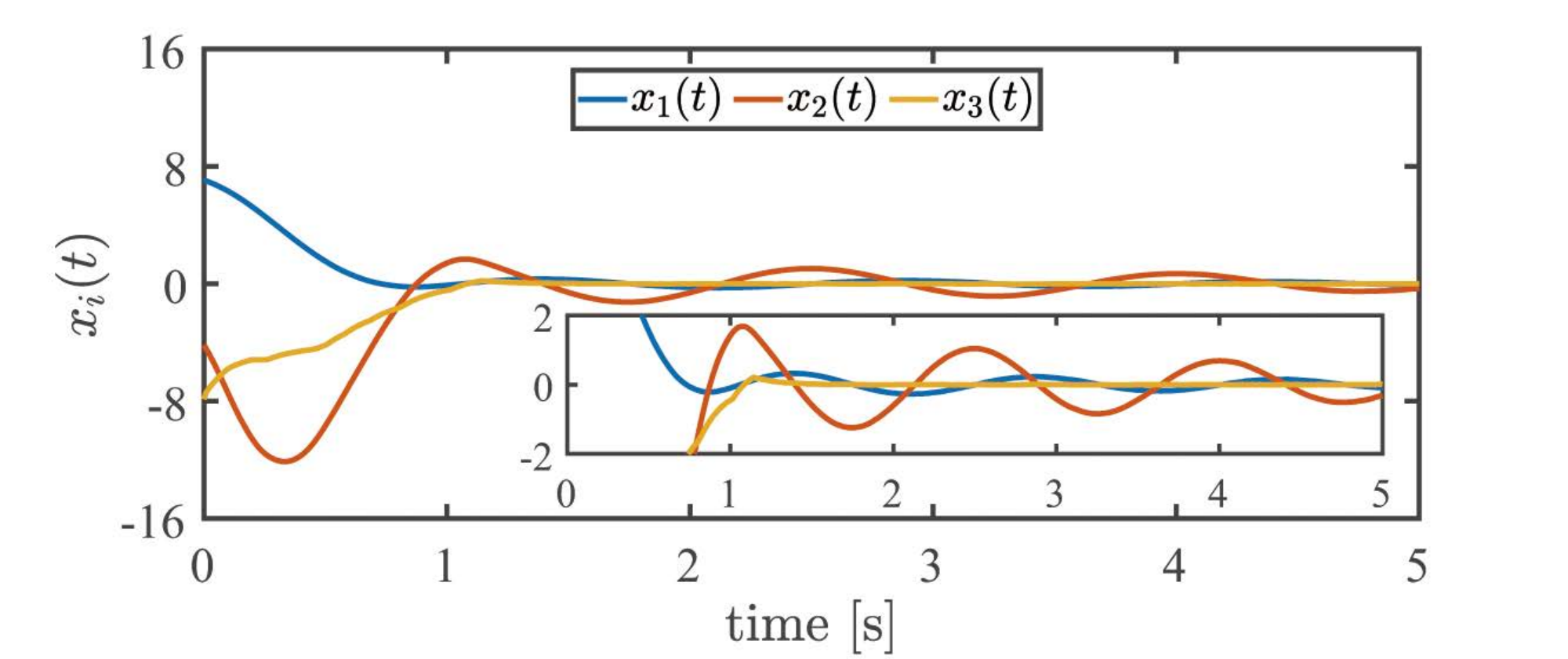}\label{fig4c}
}
\caption{The evolution of each component of the system state $x(t)$ under fixed $\bar{f}_{u}=100$ and different $\theta$ in the presence of uniform quantization.}\label{fig4}
\end{figure*}

\begin{figure*}[!htb]
\centering
\subfigure[$\theta=0.4$.]{
\includegraphics[width=5.5cm]{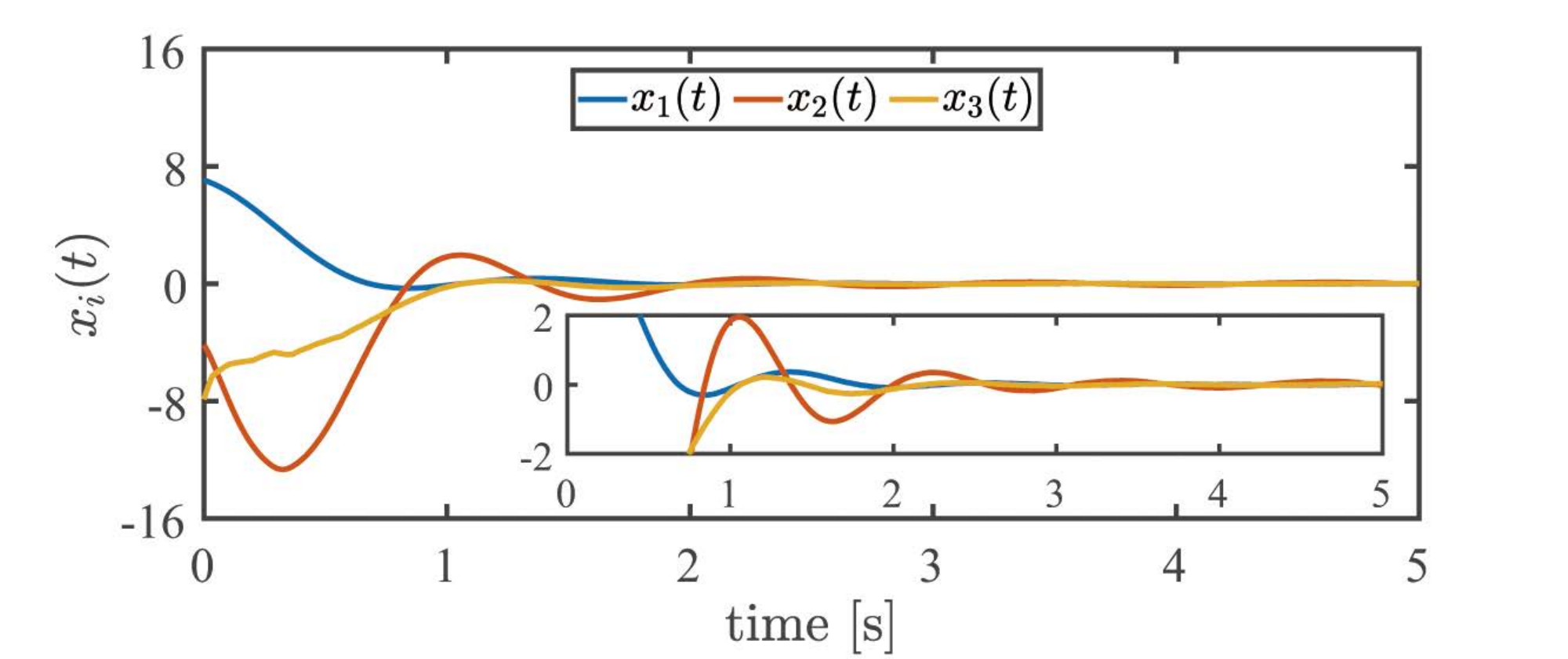}\label{fig5a}
}
\hspace{-9mm}
\subfigure[$\theta=0.6$.]{
\includegraphics[width=5.5cm]{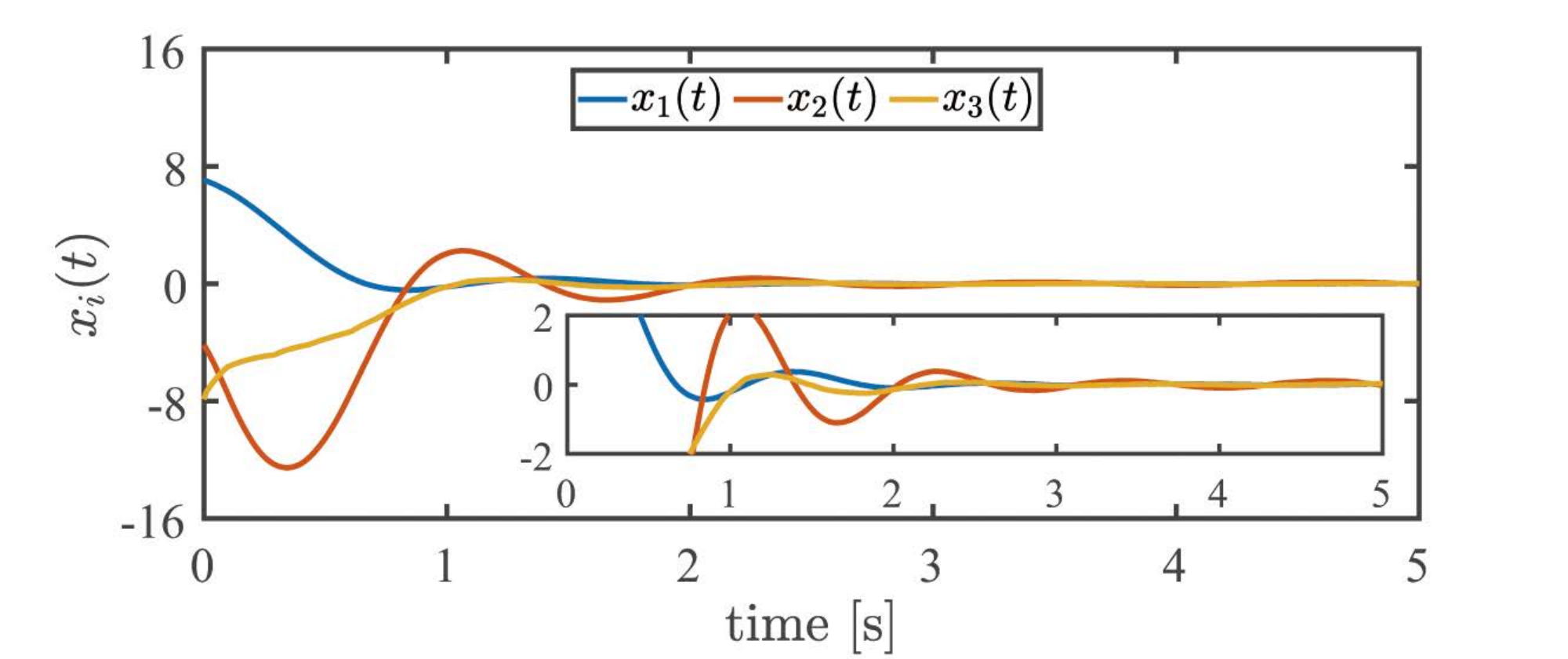}\label{fig5b}
}
\hspace{-9mm}
\subfigure[$\theta=0.8$.]{
\includegraphics[width=5.5cm]{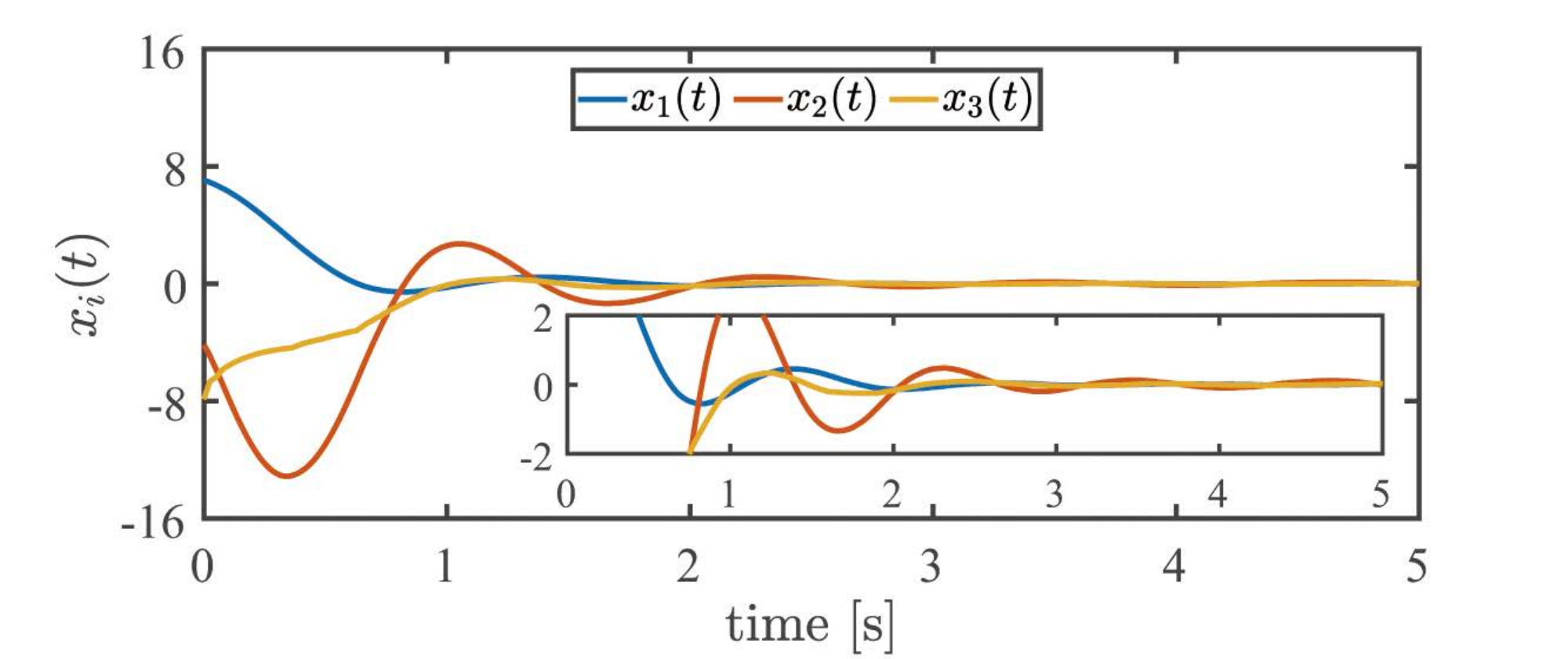}\label{fig5c}
}
\caption{The evolution of each component of the system state $x(t)$ under fixed $\bar{f}_{l}=100$ and different $\theta$ in the presence of logarithmic quantization.}\label{fig5}
\end{figure*}

To show the efficiency of the main results developed in Section~\ref{sec5} and the effect of quantizer parameter $\theta$,
the trajectories of each component of the system state $x(t)$ under fixed $\bar{d}=0.1$ and $\bar{f}_{\ast}=100$, $\ast=u, l$, and different $\theta$ are exhibited in Figs.~\ref{fig4} and~\ref{fig5}, respectively, where the initial value of $x(t)$ is the same as that in Section~\ref{sec6.2}. Figs.~\ref{fig2} and~\ref{fig4} show that the convergence error of each component of $x(t)$ increases when uniform quantization occurs, and it also increases with the increase of $\theta$. By comparing Figs.~\ref{fig2} and~\ref{fig5}, the convergence rate of each component of $x(t)$ decreases when there exists a logarithmic quantization, and it further decreases as $\theta$ increases.

\section{Conclusions}\label{sec7}
This paper presents a data-driven solution to event-triggered control problem of unknown
continuous-time linear systems in the presence of disturbances.
A dynamic ETM consisting of a dynamically updated event-triggering function is developed to determine how and when events are generated, where the designed parameters are determined only by a simple LMI.
Using state feedback at event instants, an event-based controller is proposed, where the feedback gain matrix is designed by means of a collection of data instead of precise system model.
Under the data-driven dynamic event-triggered control scheme,
the exponential ISS property of the closed-loop system is achieved, and the inter-event intervals are lower bounded by a strictly positive constant. Furthermore,
the constraint of uniform and logarithmic quantization is further studied in this paper, resulting in a quantized version of the developed data-driven dynamic event-triggered control scheme.
In the future, we would consider designing data-driven disturbance observers to handle the effects caused by unknown external disturbances.
\appendices
\section{Proof of Theorem~\ref{th1}}\label{app1}
Some useful lemmas are introduced in the following.
\begin{lemma}\label{le1}\cite{Persis2023TACEVENT}
Consider the system (\ref{eq1}) with Assumptions~\ref{as1}--\ref{as4}. Then it holds that $A+BK=(\mathcal{X}_{1}-\mathcal{D}_{0})\mathcal{G}$, where the matrix $\mathcal{G}\in\mathbb{R}^{\tau\times n}$ is any solution of
\begin{align}\label{eq25}
\left[\begin{array}{c}
K\\
\mathbf{I}\\
\end{array}\right]=
\left[\begin{array}{c}
\mathcal{U}_{0}\\
\mathcal{X}_{0}\\
\end{array}\right]\mathcal{G}.
\end{align}
\end{lemma}
\begin{lemma}\label{le2}\cite{Persis2023TACEVENT}
Suppose that Assumptions~\ref{as1}--\ref{as4} and (\ref{eq10}) hold. Then, for any $\mathcal{D}\in\mathrm{D}$, it holds that
\begin{align}\label{eq26}
(\mathcal{X}_{1}-\mathcal{D})\mathcal{Y}+
\mathcal{Y}^{T}(\mathcal{X}_{1}-\mathcal{D})^{T}+\Omega\prec0.
\end{align}
\end{lemma}

It is time to present the proof of Theorem~\ref{th1}.

\begin{proof}
The Lyapunov function is constructed as
\begin{align}\label{eq27}
V(t)=x(t)^{T}\mathcal{P}x(t)+\frac{1}{\delta}f(t),
\end{align}
which is well-defined since $\mathcal{P}=(\mathcal{X}_{0}\mathcal{Y})^{-1}\succ0$ and $f(t)\geq0$.

Substituting (\ref{eq6}) and (\ref{eq8}) into (\ref{eq1}) yields the following closed-loop system
\begin{align}\label{eq28}
\dot{x}(t)=(A+BK)x(t)+BKe(t)+d(t).
\end{align}
Note from Lemma~\ref{le1} that $A+BK=(\mathcal{X}_{1}-\mathcal{D}_{0})\mathcal{G}$ and from (\ref{eq12}) that $BK=B\mathcal{U}_{0}\mathcal{Q}=(\mathcal{X}_{1}-A\mathcal{X}_{0}-\mathcal{D}_{0})\mathcal{Q}=(\mathcal{X}_{1}-\mathcal{D}_{0})\mathcal{Q}$.
Taking the derivative of $V(t)$ along the trajectories of (\ref{eq5}) and (\ref{eq28}), one obtains
\begin{align}\label{eq29}
\dot{V}(t)=&2[(A+BK)x(t)+BKe(t)+d(t)]^{T}\mathcal{P}x(t)\notag\\
&+\frac{1}{\delta}\min\left\{-z(t)^{T}\Phi z(t), 0\right\}-\frac{1}{\delta}f(t)\notag\\
\leq&2\left[(\mathcal{X}_{1}-\mathcal{D}_{0})\mathcal{G}x(t)+(\mathcal{X}_{1}-\mathcal{D}_{0})\mathcal{Q}e(t)\right]^{T}\mathcal{P}x(t)\notag\\
&+2d(t)^{T}\mathcal{P}x(t)-\frac{1}{\delta}z(t)^{T}\Phi z(t)-\frac{1}{\delta}f(t).
\end{align}
Let $\mathcal{G}=\mathcal{Y}\mathcal{P}$, which is feasible under (\ref{eq25}) since
$\mathcal{U}_{0}\mathcal{G}=\mathcal{U}_{0}\mathcal{Y}\mathcal{P}
=\mathcal{U}_{0}\mathcal{Y}(\mathcal{X}_{0}\mathcal{Y})^{-1}=K$ and
$\mathcal{X}_{0}\mathcal{G}=\mathcal{X}_{0}\mathcal{Y}\mathcal{P}
=\mathcal{X}_{0}\mathcal{Y}(\mathcal{X}_{0}\mathcal{Y})^{-1}=\mathbf{I}$.
It follows from Lemma~\ref{le2} that
\begin{align}\label{eq30}
&2x(t)^{T}[(\mathcal{X}_{1}-\mathcal{D}_{0})\mathcal{G}]^{T}\mathcal{P}x(t)\notag\\
=&x(t)^{T}\mathcal{P}(\mathcal{X}_{1}-\mathcal{D}_{0})\mathcal{G}x(t)+x(t)^{T}\mathcal{G}^{T}(\mathcal{X}_{1}-\mathcal{D}_{0})^{T}\mathcal{P}x(t)\notag\\
=&x(t)^{T}\mathcal{P}\left[(\mathcal{X}_{1}-\mathcal{D}_{0})\mathcal{Y}+\mathcal{Y}^{T}(\mathcal{X}_{1}-\mathcal{D}_{0})^{T}\right]\mathcal{P}x(t)\notag\\
\leq&-x(t)^{T}\mathcal{P}\Omega\mathcal{P}x(t).
\end{align}
Given a constant $\iota\geq\frac{8\lambda_{M}(\mathcal{P}^{2})}{\lambda_{m}(\mathcal{P}\Omega\mathcal{P})}$. By means of Young's inequality, one has
\begin{align}\label{eq31}
2d(t)^{T}\mathcal{P}x(t)\leq&\iota d(t)^{T}d(t)+\frac{1}{\iota}x(t)^{T}\mathcal{P}^{2}x(t)\notag\\
\leq&\iota d(t)^{T}d(t)+\frac{1}{8}x(t)^{T}\mathcal{P}\Omega\mathcal{P}x(t).
\end{align}
It can be obtained from (\ref{eq29})--(\ref{eq31}) that
\begin{align}\label{eq32}
\dot{V}(t)\leq&-\frac{3}{4}x(t)^{T}\mathcal{P}\Omega\mathcal{P}x(t)+z(t)^{T}\Lambda z(t)\notag\\
&-\frac{1}{\delta}z(t)^{T}\Phi z(t)-\frac{1}{\delta}f(t)+\iota d(t)^{T}d(t),
\end{align}
where
\begin{align}\label{eq33}
\Lambda=\left[\begin{array}{cc}
-\frac{1}{8}\mathcal{P}\Omega\mathcal{P}&\mathcal{P}(\mathcal{X}_{1}-\mathcal{D}_{0})\mathcal{Q}\\
\star&\mathbf{0}\\
\end{array}\right].
\end{align}
By the Schur complement, the LMI given in (\ref{eq14}) is equivalent to that
\begin{align}\label{eq34}
\delta\left[\begin{array}{cc}
-\frac{1}{8}\mathcal{P}\Omega\mathcal{P}&\mathcal{P}\mathcal{X}_{1}\mathcal{Q}\\
\star&\mathbf{0}\\
\end{array}\right]
-&\Phi
+\gamma\left[\begin{array}{c}
\mathbf{0}\\
\mathcal{Q}^{T}\\
\end{array}\right]
\left[\begin{array}{cc}
\mathbf{0}&\mathcal{Q}\\
\end{array}\right]\notag\\
+\frac{1}{\gamma}
\left[\begin{array}{c}
\delta\mathcal{P}\Delta\\
\mathbf{0}\\
\end{array}\right]&
\left[\begin{array}{cc}
\delta\Delta^{T}\mathcal{P}&\mathbf{0}\\
\end{array}\right]\preceq0.
\end{align}
Let
\begin{align}\label{eq35}
\Theta=\left[\begin{array}{c}
\mathbf{0}\\
\mathcal{Q}^{T}\\
\end{array}\right],
\Upsilon=\left[\begin{array}{c}
-\delta\mathcal{P}\\
\mathbf{0}\\
\end{array}\right]^{T}.
\end{align}
Note that
\begin{align}\label{eq36}
\left(\sqrt{\gamma}\Theta-\sqrt{\frac{1}{\gamma}}\Upsilon^{T}\mathcal{D}_{0}\right)
\left(\sqrt{\gamma}\Theta-\sqrt{\frac{1}{\gamma}}\Upsilon^{T}\mathcal{D}_{0}\right)^{T}\succeq0,
\end{align}
which indicates that
\begin{align}\label{eq37}
\Theta\mathcal{D}_{0}^{T}\Upsilon+\Upsilon^{T}\mathcal{D}_{0}\Theta^{T}
\preceq&\gamma\Theta\Theta^{T}+\frac{1}{\gamma}\Upsilon^{T}\mathcal{D}_{0}\mathcal{D}_{0}^{T}\Upsilon\notag\\
\preceq&\gamma\Theta\Theta^{T}+\frac{1}{\gamma}\Upsilon^{T}\Delta\Delta^{T}\Upsilon,
\end{align}
where Assumption~\ref{as4} is used.
Substituting (\ref{eq35}) into (\ref{eq37}) and combining (\ref{eq34}) yields that
\begin{align}\label{eq38}
\delta\left[\begin{array}{cc}
-\frac{1}{8}\mathcal{P}\Omega\mathcal{P}&\mathcal{P}\mathcal{X}_{1}\mathcal{Q}\\
\star&\mathbf{0}\\
\end{array}\right]
-&\Phi
+\left[\begin{array}{c}
\mathbf{0}\\
\mathcal{Q}^{T}\\
\end{array}\right]\mathcal{D}_{0}^{T}
\left[\begin{array}{c}
-\delta\mathcal{P}\\
\mathbf{0}\\
\end{array}\right]^{T}\notag\\
+
\left[\begin{array}{c}
-\delta\mathcal{P}\\
\mathbf{0}\\
\end{array}\right]&\mathcal{D}_{0}
\left[\begin{array}{c}
\mathbf{0}\\
\mathcal{Q}^{T}\\
\end{array}\right]^{T}\preceq0,
\end{align}
implying that $\delta\Lambda-\Phi\preceq0$. It thus holds that
\begin{align}\label{eq39}
z(t)^{T}\Lambda z(t)-\frac{1}{\delta}z(t)^{T}\Phi z(t)\leq0.
\end{align}
Combining (\ref{eq32}) and (\ref{eq39}) yields that
\begin{align}\label{eq40}
\dot{V}(t)\leq&-\frac{3}{4}x(t)^{T}\mathcal{P}\Omega\mathcal{P}x(t)-\frac{1}{\delta}f(t)+\iota d(t)^{T}d(t)\notag\\
\leq&-\bar{\lambda}V(t)+\iota d(t)^{T}d(t),
\end{align}
where $\bar{\lambda}=\min\{\frac{3\lambda_{m}(\mathcal{P}\Omega\mathcal{P})}{4\lambda_{M}(\mathcal{P})}, 1\}$.
It can be concluded from (\ref{eq40}) that the system (\ref{eq1}) is with exponentially ISS property with respect to the disturbance, implying that Theorem~\ref{th1}~(i) holds.

From (\ref{eq6}), it is easy to verify that $\|e(t)\|$ is bounded. It is assumed that $\|e(t)\|\leq\bar{e}$, where the constant $\bar{e}>0$.
From (\ref{eq5}), one obtains
\begin{align}\label{eq41}
\dot{f}(t)\geq&-\beta\bar{e}^{2}-\bar{f}, f(t_{k}^{+})=\bar{f}.
\end{align}
By the comparison lemma, $f(t)\geq g(t)$ for $t\in[t_{k}, t_{k+1})$, with $g(t)$ being the solution of
\begin{align}\label{eq42}
\dot{g}(t)=-\beta\bar{e}^{2}-\bar{f}, g(t_{k}^{+})=\bar{f}.
\end{align}
Solving (\ref{eq42}) obtains that
\begin{align}\label{eq43}
t-t_{k}=\frac{\bar{f}}{\beta\bar{e}^{2}+\bar{f}}-\frac{g(t)}{\beta\bar{e}^{2}+\bar{f}}.
\end{align}
Since $f(t)\geq g(t)$, the inter-event time $t_{k+1}-t_{k}$ is lower bounded by the time that it takes for $g(t)$ to decrease from $\bar{f}$ to $0$. It thus follows from (\ref{eq43}) that
\begin{align}\label{eq44}
t_{k+1}-t_{k}\geq\frac{\bar{f}}{\beta\bar{e}^{2}+\bar{f}}.
\end{align}
It is concluded from (\ref{eq44}) that the lower bound of the inter-event intervals $t_{k+1}-t_{k}, k\in\mathbb{N}$ is strictly positive for any time, which validates Theorem~\ref{th1}~(ii).
\end{proof}
\section{Proof of Theorem~\ref{th2}}\label{app2}
Under the quantization effect, the continuity of the vector field of the closed-loop system may not hold any more.
By using the non-smooth analysis, the Filippov solution of the system subject to quantization is introduced as follows \cite{Filippov1988}.

Given the system $\dot{y}(t)=f(y(t))$, where $f: \mathbb{R}^{n}\rightarrow\mathbb{R}^{n}$ is a measurable and essentially locally bounded function.
A Filippov set-valued map $F[f](y(t)): \mathbb{R}^{n}\rightarrow\mathcal{B}(\mathbb{R}^{n})$ is defined by
$F[f](y(t))=\cap_{r>0}\cap_{\mu(\mathrm{S})=0}\overline{\mathrm{co}}\{f(\mathbf{B}(y(t), r)\backslash\mathrm{S})\}$, where $\mathcal{B}(\mathbb{R}^{n})$ denotes the collection of subsets of $\mathbb{R}^{n}$, $\mathbf{B}(y(t), r)$ is
an open ball centered at $y(t)$ with radius $r$, $\overline{\mathrm{co}}$ denotes the convex closure, $\mu(\mathrm{S})$ denotes the Lebesgue measure of the set $\mathrm{S}$.
By definition,
Filippov solutions are absolutely continuous curves, which satisfy the differential inclusion of the form
$\dot{y}(t)\in F[f](y(t))$ for almost all $t\geq0$.

The proof of Theorem~\ref{th2} is presented as follows.

\begin{proof}
We Adopt the following Lyapunov function
\begin{align}\label{eq45}
V_{\ast}(t)=x(t)^{T}\mathcal{P}x(t)+\frac{1}{\delta}f_{\ast}(t),
\end{align}
where $\ast=u, l$.

Recall that $e_{\ast}(t)=q_{\ast}(x(t_{k}))-q_{\ast}(x(t))$ and $\epsilon_{\ast}(t)=q_{\ast}(x(t))-x(t)$. Let $\tilde{e}_{\ast}(t)=F[q_{\ast}](x(t_{k}))-F[q_{\ast}](x(t))$, $\tilde{\epsilon}_{\ast}(t)=F[q_{\ast}](x(t))-x(t)$.
It follows from (\ref{eq1}), (\ref{eq22}) and (\ref{eq24}) that
\begin{align}\label{eq46}
\dot{x}(t)=&(A+BK)x(t)+BKe_{\ast}(t)+BK\epsilon_{\ast}(t)+d(t)\notag\\
\in&(A+BK)x(t)+BK\tilde{e}_{\ast}(t)+BK\tilde{\epsilon}_{\ast}(t)+d(t).
\end{align}
Similarly as in deriving (\ref{eq32}), the derivative of $V_{\ast}(t)$ along the trajectories of (\ref{eq21}) and (\ref{eq46}) is calculated as
$\dot{V}_{\ast}(t)\in\dot{\tilde{V}}_{\ast}(t)$, for a.e. $t\geq0$, where
\begin{align}\label{eq47}
\dot{\tilde{V}}_{\ast}(t)\subseteq&-\frac{3}{4}x(t)^{T}\mathcal{P}\Omega\mathcal{P}x(t)+\hat{z}_{\ast}(t)^{T}\Lambda \hat{z}_{\ast}(t)\notag\\
&-\frac{1}{\delta}\tilde{z}_{\ast}(t)^{T}\Phi_{\ast}\tilde{z}_{\ast}(t)-\frac{1}{\delta}f_{\ast}(t)+\iota d(t)^{T}d(t)\notag\\
&+2\tilde{\epsilon}_{\ast}(t)^{T}\mathcal{Q}^{T}(\mathcal{X}_{1}-\mathcal{D}_{0})^{T}\mathcal{P}x(t),
\end{align}
where $\hat{z}_{\ast}(t)=[x(t)^{T}\ \ \tilde{e}_{\ast}(t)^{T}]^{T}$, $\tilde{z}_{\ast}(t)=[F[q_{\ast}](x(t))^{T}\ \ \tilde{e}_{\ast}(t)^{T}]^{T}$, and $\Lambda$ remains the same as that in (\ref{eq33}).
Under Assumption~\ref{as4}, $\|\mathcal{D}_{0}\|^{2}\leq\|\Delta\|^{2}$, which leads to that $\|(\mathcal{X}_{1}-\mathcal{D}_{0})\mathcal{Q}\|^{2}\leq\|\mathcal{Q}\|^{2}\|(\mathcal{X}_{1}-\mathcal{D}_{0})\|^{2}\leq2\|\mathcal{Q}\|^{2}(\|\mathcal{X}_{1}\|^{2}+\|\mathcal{D}_{0}\|^{2})\leq2\|\mathcal{Q}\|^{2}(\|\mathcal{X}_{1}\|^{2}+\|\Delta\|^{2})$.
Using Young's inequality, one has
\begin{align}\label{eq48}
&2\tilde{\epsilon}_{\ast}(t)^{T}\mathcal{Q}^{T}(\mathcal{X}_{1}-\mathcal{D}_{0})^{T}\mathcal{P}x(t)\notag\\
\leq&\iota\|(\mathcal{X}_{1}-\mathcal{D}_{0})\mathcal{Q}\|^{2}\tilde{\epsilon}_{\ast}(t)^{T}\tilde{\epsilon}_{\ast}(t)+\frac{1}{\iota}x(t)^{T}\mathcal{P}^{2}x(t)\notag\\
\leq&2\iota\|\mathcal{Q}\|^{2}\left(\|\mathcal{X}_{1}\|^{2}+\|\Delta\|^{2}\right)\tilde{\epsilon}_{\ast}(t)^{T}\tilde{\epsilon}_{\ast}(t)
+\frac{1}{8}x(t)^{T}\mathcal{P}\Omega\mathcal{P}x(t),
\end{align}
where $\iota\geq\frac{8\lambda_{M}(\mathcal{P}^{2})}{\lambda_{m}(\mathcal{P}\Omega\mathcal{P})}$ is used.

In the following, we separate the proof with two cases according to the uniform quantization $(\ast=u)$ and logarithmic quantization $(\ast=l)$.

Case I: $\ast=u$. Note that
$F[q_{u}](x(t))^{T}F[q_{u}](x(t))=(x(t)+\tilde{\epsilon}_{u}(t))^{T}(x(t)+\tilde{\epsilon}_{u}(t))
\leq2(x(t)^{T}x(t)+\tilde{\epsilon}_{u}(t)^{T}\tilde{\epsilon}_{u}(t))$.
It thus follows from (\ref{eq23}) that
\begin{align}\label{eq49}
-\frac{1}{\delta}\tilde{z}_{u}(t)^{T}\Phi_{u}\tilde{z}_{u}(t)
\leq&
-\frac{1}{\delta}\hat{z}_{u}(t)^{T}\Phi\hat{z}_{u}(t)+\frac{\alpha}{\delta}\tilde{\epsilon}_{u}(t)^{T}\tilde{\epsilon}_{u}(t)\notag\\
\leq&
-\frac{1}{\delta}\hat{z}_{u}(t)^{T}\Phi\hat{z}_{u}(t)+\frac{\alpha n\theta^{2}}{4\delta},
\end{align}
where $\Phi$ specified in (\ref{eq7}) and $\|\tilde{\epsilon}_{u}(t)\|\leq\frac{\sqrt{n}\theta}{2}$ are used.
Similarly as in obtaining (\ref{eq39}), one has
\begin{align}\label{eq50}
\hat{z}_{u}(t)^{T}\Lambda\hat{z}_{u}(t)-\frac{1}{\delta}\hat{z}_{u}(t)^{T}\Phi\hat{z}_{u}(t)\leq0.
\end{align}
Applying $\|\tilde{\epsilon}_{u}(t)\|\leq\frac{\sqrt{n}\theta}{2}$ again, from (\ref{eq48}), one has
\begin{align}\label{eq51}
&2\tilde{\epsilon}_{u}(t)^{T}\mathcal{Q}^{T}(\mathcal{X}_{1}-\mathcal{D}_{0})^{T}\mathcal{P}x(t)\notag\\
\leq&\frac{1}{8}x(t)^{T}\mathcal{P}\Omega\mathcal{P}x(t)+\frac{\iota n\theta^{2}\|\mathcal{Q}\|^{2}\left(\|\mathcal{X}_{1}\|^{2}+\|\Delta\|^{2}\right)}{2}.
\end{align}
It can be derived from (\ref{eq47}), (\ref{eq49})--(\ref{eq51}) that
\begin{align}\label{eq52}
\dot{\tilde{V}}_{u}(t)\subseteq&-\frac{5}{8}x(t)^{T}\mathcal{P}\Omega\mathcal{P}x(t)-\frac{1}{\delta}f_{u}(t)
+\iota d(t)^{T}d(t)+\bar{\epsilon}_{u}\notag\\
\subseteq&-\bar{\lambda}_{u}V_{u}(t)+\iota d(t)^{T}d(t)+\bar{\epsilon}_{u},
\end{align}
where $\bar{\lambda}_{u}=\min\{\frac{5\lambda_{m}(\mathcal{P}\Omega\mathcal{P})}{8\lambda_{M}(\mathcal{P})}, 1\}$, $\bar{\epsilon}_{u}=\frac{\iota n\theta^{2}\|\mathcal{Q}\|^{2}(\|\mathcal{X}_{1}\|^{2}+\|\Delta\|^{2})}{2}+\frac{\alpha n\theta^{2}}{4\delta}$.
which means that the system (\ref{eq1}) subject to uniform quantization is with practical exponential
ISS property.

Case II: $\ast=l$. Note that $\|\tilde{\epsilon}_{l}(t)\|\leq(e^{\frac{\theta}{2}}-1)\|x(t)\|$. Combining
$\|\tilde{\epsilon}_{l}(t)\|\geq\|F[q_{l}](x(t))\|-\|x(t)\|$ leads to that $\|F[q_{l}](x(t))\|\leq e^{\frac{\theta}{2}}\|x(t)\|$.
Therefore, it holds that $e^{-\theta}F[q_{l}](x(t))^{T}F[q_{l}](x(t))\leq x(t)^{T}x(t)$, which indicates that
\begin{align}\label{eq53}
-\frac{1}{\delta}\tilde{z}_{l}(t)^{T}\Phi_{l}\tilde{z}_{l}(t)\leq-\frac{1}{\delta}\hat{z}_{l}(t)^{T}\Phi\hat{z}_{l}(t),
\end{align}
with $\Phi$ given in (\ref{eq7}).
Similarly as in deriving (\ref{eq39}), one has
\begin{align}\label{eq54}
\hat{z}_{l}(t)^{T}\Lambda\hat{z}_{l}(t)-\frac{1}{\delta}\hat{z}_{l}(t)^{T}\Phi\hat{z}_{l}(t)\leq0.
\end{align}
Choose $\theta\leq2\ln\left(1+\frac{1}{4}\sqrt{\frac{\lambda_{m}(\mathcal{P}\Omega\mathcal{P})}{\iota\|\mathcal{Q}\|^{2}(\|\mathcal{X}_{1}\|^{2}+\|\Delta\|^{2})}}\right)$.
It can be obtained that
\begin{align}\label{eq55}
&2\iota\|\mathcal{Q}\|^{2}\left(\|\mathcal{X}_{1}\|^{2}+\|\Delta\|^{2}\right)\tilde{\epsilon}_{l}(t)^{T}\tilde{\epsilon}_{l}(t)\notag\\
\leq&2\iota\|\mathcal{Q}\|^{2}\left(\|\mathcal{X}_{1}\|^{2}+\|\Delta\|^{2}\right)\left(e^{\frac{\theta}{2}}-1\right)^{2}x(t)^{T}x(t)\notag\\
\leq&\frac{1}{8}x(t)^{T}\mathcal{P}\Omega\mathcal{P}x(t),
\end{align}
where $\|\tilde{\epsilon}_{l}(t)\|\leq(e^{\frac{\theta}{2}}-1)\|x(t)\|$ is utilized.
From (\ref{eq48}) and (\ref{eq55}), one has
\begin{align}\label{eq56}
2\tilde{\epsilon}_{l}(t)^{T}\mathcal{Q}^{T}(\mathcal{X}_{1}-\mathcal{D}_{0})^{T}\mathcal{P}x(t)
\leq\frac{1}{4}x(t)^{T}\mathcal{P}\Omega\mathcal{P}x(t).
\end{align}
Combining (\ref{eq47}), (\ref{eq53}), (\ref{eq54}) and (\ref{eq56}), one obtains
\begin{align}\label{eq57}
\dot{\tilde{V}}_{l}(t)\subseteq&-\frac{1}{2}x(t)^{T}\mathcal{P}\Omega\mathcal{P}x(t)-\frac{1}{\delta}f_{l}(t)+\iota d(t)^{T}d(t)\notag\\
\subseteq&-\bar{\lambda}_{l}V_{l}(t)+\iota d(t)^{T}d(t),
\end{align}
where $\bar{\lambda}_{l}=\min\{\frac{\lambda_{m}(\mathcal{P}\Omega\mathcal{P})}{2\lambda_{M}(\mathcal{P})}, 1\}$.
It can be concluded from (\ref{eq57}) that the system (\ref{eq1}) in the presence of logarithmic quantization features the exponential ISS property with respect to disturbance.

Summarizing Cases I and II yields that Theorem~\ref{th2}~(i) and~(ii) hold. Moreover, adopting the same analysis as in deriving (\ref{eq44}), one derives
\begin{align}\label{eq58}
t_{k+1}-t_{k}\geq\frac{\bar{f}_{\ast}}{\beta\bar{e}_{\ast}^{2}+\bar{f}_{\ast}},
\end{align}
where the positive constant $\bar{e}_{\ast}$ represents the upper bound of $\|\tilde{e}_{\ast}(t)\|$, $\ast=u, l$.
Therefore, Theorem~\ref{th2}~(iii) holds.
\end{proof}
\bibliographystyle{IEEEtran}
\bibliography{ref}

\begin{thebibliography}{10}
\providecommand{\url}[1]{#1}
\csname url@samestyle\endcsname
\providecommand{\newblock}{\relax}
\providecommand{\bibinfo}[2]{#2}
\providecommand{\BIBentrySTDinterwordspacing}{\spaceskip=0pt\relax}
\providecommand{\BIBentryALTinterwordstretchfactor}{4}
\providecommand{\BIBentryALTinterwordspacing}{\spaceskip=\fontdimen2\font plus
\BIBentryALTinterwordstretchfactor\fontdimen3\font minus
  \fontdimen4\font\relax}
\providecommand{\BIBforeignlanguage}[2]{{%
\expandafter\ifx\csname l@#1\endcsname\relax
\typeout{** WARNING: IEEEtran.bst: No hyphenation pattern has been}%
\typeout{** loaded for the language `#1'. Using the pattern for}%
\typeout{** the default language instead.}%
\else
\language=\csname l@#1\endcsname
\fi
#2}}
\providecommand{\BIBdecl}{\relax}
\BIBdecl

\bibitem{Hespanha2007IEEEA}
J.~P. Hespanha, P.~Naghshtabrizi, and Y.~Xu, ``A survey of recent results in
  networked control systems,'' \emph{Proceedings of the IEEE}, vol.~95, no.~1,
  pp. 138--162, 2007.

\bibitem{Zhang2016TIISUR}
X.-M. Zhang, Q.-L. Han, and X.~Yu, ``Survey on recent advances in networked
  control systems,'' \emph{IEEE Transactions on Industrial Informatics},
  vol.~12, no.~5, pp. 1740--1752, 2016.

\bibitem{Karl1999IFACA}
K.-E. \.{A}arz\'{e}n, ``A simple event-based pid controller,'' in \emph{Proc.
  of the 14th IFAC World Congress}, vol.~32, no.~2, 1999, pp. 8687--8692.

\bibitem{Nowzari2019AUTOEVE}
C.~Nowzari, E.~Garcia, and J.~Cort\'{e}s, ``Event-triggered communication and
  control of networked systems for multi-agent consensus,'' \emph{Automatica},
  vol. 105, pp. 1--27, 2019.

\bibitem{Ge2020SMCDYN}
X.~Ge, Q.-L. Han, L.~Ding, Y.-L. Wang, and X.-M. Zhang, ``Dynamic
  event-triggered distributed coordination control and its applications: A
  survey of trends and techniques,'' \emph{IEEE Transactions on Systems, Man,
  and Cybernetics: Systems}, vol.~50, no.~9, pp. 3112--3125, 2020.

\bibitem{Abdelrahim2016tacstabilization}
M.~Abdelrahim, R.~Postoyan, J.~Daafouz, and D.~Ne\v{s}i\'{c}, ``Stabilization
  of nonlinear systems using event-triggered output feedback controllers,''
  \emph{IEEE Transactions on Automatic Control}, vol.~61, no.~9, pp.
  2682--2687, 2016.

\bibitem{Lunze2010AUTOA}
J.~Lunze and D.~Lehmann, ``A state-feedback approach to event-based control,''
  \emph{Automatica}, vol.~46, no.~1, pp. 211--215, 2010.

\bibitem{Zhang2014AUTOEVE}
J.~Zhang and G.~Feng, ``Event-driven observer-based output feedback control for
  linear systems,'' \emph{Automatica}, vol.~50, no.~7, pp. 1852--1859, 2014.

\bibitem{Tallapragada2014tacdecentralized}
P.~Tallapragada and N.~Chopra, ``Decentralized event-triggering for control of
  nonlinear systems,'' \emph{IEEE Transactions on Automatic Control}, vol.~59,
  no.~12, pp. 3312--3324, 2014.

\bibitem{Girard2015TACDYN}
A.~Girard, ``Dynamic triggering mechanisms for event-triggered control,''
  \emph{IEEE Transactions on Automatic Control}, vol.~60, no.~7, pp.
  1992--1997, 2015.

\bibitem{Persis2020TACFOR}
C.~De~Persis and P.~Tesi, ``Formulas for data-driven control: Stabilization,
  optimality, and robustness,'' \emph{IEEE Transactions on Automatic Control},
  vol.~65, no.~3, pp. 909--924, 2020.

\bibitem{Berberich2021TACDATA}
J.~Berberich, J.~K\"{o}hler, M.~A. M\"{u}ller, and F.~Allg\"{o}wer,
  ``Data-driven model predictive control with stability and robustness
  guarantees,'' \emph{IEEE Transactions on Automatic Control}, vol.~66, no.~4,
  pp. 1702--1717, 2021.

\bibitem{Bisoffi2022AUTODAT}
A.~Bisoffi, C.~{De Persis}, and P.~Tesi, ``Data-driven control via {P}etersen's
  lemma,'' \emph{Automatica}, vol. 145, p. 110537, 2022.

\bibitem{Joscha2023TACROB}
J.~Bongard, J.~Berberich, J.~Koehler, and F.~Allg\"{o}wer, ``Robust stability
  analysis of a simple data-driven model predictive control approach,''
  \emph{IEEE Transactions on Automatic Control}, vol.~68, no.~5, pp.
  2625--2637, 2023.

\bibitem{Liu2023TACDATA}
W.~Liu, J.~Sun, G.~Wang, F.~Bullo, and J.~Chen, ``Data-driven resilient
  predictive control under denial-of-service,'' \emph{IEEE Transactions on
  Automatic Control}, vol.~68, no.~8, pp. 4722--4737, 2023.

\bibitem{Mohammad2023TACDAT}
M.~Alsalti, V.~G. Lopez, J.~Berberich, F.~Allg\"{o}wer, and M.~A. M\'{u}ller,
  ``Data-based control of feedback linearizable systems,'' \emph{IEEE
  Transactions on Automatic Control}, vol.~68, no.~11, pp. 7014--7021, 2023.

\bibitem{Cordovil2022IJRNCLEARNING}
L.~A.~Q. Cordovil~Jr, P.~H.~S. Coutinho, I.~Bessa, M.~L.~C. Peixoto, and R.~M.
  Palhares, ``Learning event-triggered control based on evolving data-driven
  fuzzy granular models,'' \emph{International Journal of Robust and Nonlinear
  Control}, vol.~32, no.~5, pp. 2805--2827, 2022.

\bibitem{Digge2022ECCDATA}
V.~Digge and R.~Pasumarthy, ``Data-driven event-triggered control for
  discrete-time {LTI} systems,'' in \emph{Proc. of the 2022 European Control
  Conference (ECC)}, 2022, pp. 1355--1360.

\bibitem{Wang2022arXivDATA}
X.~Wang, J.~Berberich, J.~Sun, G.~Wang, F.~Allgöwer, and J.~Chen,
  ``Model-based and data-driven control of event- and self-triggered
  discrete-time linear systems,'' \emph{IEEE Transactions on Cybernetics},
  vol.~53, no.~9, pp. 6066--6079, 2023.

\bibitem{Wei2023IJRNCDATA}
Z.-J. Wei, C.-Y. Xu, K.-Z. Liu, W.-L. Qi, and X.-M. Sun, ``Data-driven analysis
  and control of periodic event-triggered continuous-time systems,''
  \emph{International Journal of Robust and Nonlinear Control}, vol.~33,
  no.~13, pp. 7951--7967, 2023.

\bibitem{Qi2023TIEDATA}
W.-L. Qi, K.-Z. Liu, R.~Wang, and X.-M. Sun, ``Data-driven $\mathcal
  {L}_{2}$-stability analysis for dynamic event-triggered networked control
  systems: A hybrid system approach,'' \emph{IEEE Transactions on Industrial
  Electronics}, vol.~70, no.~6, pp. 6151--6158, 2023.

\bibitem{Wang2021arXivDATA}
X.~Wang, J.~Sun, J.~Berberich, G.~Wang, F.~Allg\"{o}wer, and J.~Chen,
  ``Data-driven control of dynamic event-triggered systems with delays,''
  \emph{International Journal of Robust and Nonlinear Control}, vol.~33,
  no.~12, pp. 7071--7093, 2023.

\bibitem{Persis2023TACEVENT}
C.~D. Persis, R.~Postoyan, and P.~Tesi, ``Event-triggered control from data,''
  \emph{IEEE Transactions on Automatic Control}, pp. 1--16, 2023.

\bibitem{Berberich2021arXivdata}
J.~Berberich, S.~Wildhagen, M.~Hertneck, and F.~Allg\"{o}wer, ``Data-driven
  analysis and control of continuous-time systems under aperiodic sampling,''
  \emph{arXiv preprint arXiv:2011.09221}, 2021.

\bibitem{Guo2013AUTOCON}
M.~Guo and D.~V. Dimarogonas, ``Consensus with quantized relative state
  measurements,'' \emph{Automatica}, vol.~49, no.~8, pp. 2531--2537, 2013.

\bibitem{heck1989jgcdapplication}
B.~S. Heck and A.~A. Ferri, ``Application of output feedback to variable
  structure systems,'' \emph{Journal of Guidance, Control, and Dynamics},
  vol.~12, no.~6, pp. 932--935, 1989.

\bibitem{Filippov1988}
A.~F. Filippov and F.~M. Arscott, \emph{Differential equations with
  discontinuous righthand sides}.\hskip 1em plus 0.5em minus 0.4em\relax
  Springer, 1988.

\end{thebibliography}
\end{document}